\documentclass[10pt,aps,preprint,prd,showpacs,nofootinbib,preprintnumbers,a4paper,superscriptaddress]{revtex4-1}
\pdfoutput=1
\usepackage{amsmath}
\usepackage{amssymb}
\usepackage[pdftex]{graphicx}
\usepackage{epstopdf}
\usepackage[caption=false]{subfig}
\usepackage[english]{babel}
\usepackage{amsfonts}
\usepackage{bbold}
\usepackage{bm}
\usepackage{color}
\usepackage{xcolor}
\usepackage{dcolumn}
\usepackage[colorlinks]{hyperref}
\usepackage{cancel}
\usepackage{units}
\usepackage[normalem]{ulem}
\usepackage{slashed}
\usepackage{feynmp}
\usepackage{ulem}
\hypersetup{
pdfauthor={},
pdftitle={strongly coupled hidden sector}
}

\newcommand{\diag}{\ensuremath{\mathrm{diag}}}

\newcommand{\GeV}{\ensuremath{\,\mathrm{GeV}}}

\newcommand{\SU}[1]{\ensuremath{\mathrm{SU}(#1)}}

\newcommand{\nn}{\nonumber}

\newcommand{\be}{\begin{eqnarray}}
\newcommand{\ee}{\end{eqnarray}}

\begin{document}
\allowdisplaybreaks[1]

\title{Electroweak and Conformal Symmetry Breaking \protect\\  by a Strongly Coupled Hidden Sector \vspace{7mm}}

\author{Martin Holthausen}
\email{martin.holthausen@mpi-hd.mpg.de}
\affiliation{Max-Planck-Institut f\"{u}r Kernphysik, 69117 Heidelberg, Germany}

\author{Jisuke Kubo}
\email{jik@hep.s.kanazawa-u.ac.jp}
\affiliation{Institute for Theoretical Physics, Kanazawa University, Kanazawa 920-1192, Japan  \vspace{10mm}}

\author{Kher Sham Lim}
\email{khersham.lim@mpi-hd.mpg.de}
\affiliation{Max-Planck-Institut f\"{u}r Kernphysik, 69117 Heidelberg, Germany}

\author{ Manfred Lindner  \vspace{2mm}}
\email{lindner@mpi-hd.mpg.de}
\affiliation{Max-Planck-Institut f\"{u}r Kernphysik, 69117 Heidelberg, Germany}

\preprint{KANAZAWA-10-13}
\vspace*{1cm}
\begin{abstract}
The LHC and other experiments show so far no sign of new physics and long-held beliefs about naturalness should be critically reexamined. We discuss therefore in this paper a model with a combined breaking of conformal and electroweak symmetry by a strongly coupled hidden sector. Even though the conformal symmetry is anomalous, this may still provide an explanation of the smallness of electroweak scale compared to the Planck scale. Specifically we start from a classically conformal model, in which a strongly coupled hidden sector undergoes spontaneous chiral symmetry breaking. A coupling via a real scalar field transmits the breaking scale to the Standard Model Higgs and triggers electroweak symmetry breaking. The model contains dark matter candidates in the form of dark pions, whose stability is being guaranteed by the flavor symmetry of hidden quark sector. We study its relic abundance and direct detection prospects with the Nambu-Jona-Lasinio method and discuss the phase transition in the dark sector as 
well as in 
the electroweak sector.
\end{abstract}

\maketitle

\section{Introduction \label{sec:introduction}}
At the present level of our understanding, nature has (at least) three fundamental scales: the Planck scale $M_{Pl}$, where gravitational interactions become strong, the QCD scale $\Lambda_{QCD}$ at which the interactions of quantum chromodynamics (QCD) grow strong and the electroweak (EW) scale $v=246 \GeV$, around where - in the absence of the Higgs - the longitudinal gauge boson interactions would have become strong.
Of the two dimensionless scale ratios one can form, the smallness of the ratio $\Lambda_{QCD}/M_{Pl}\sim 10^{-19}$ is naturally explained as a consequence of the logarithmic running of the QCD gauge coupling. The hierarchy of the EW scale to the Planck scale $v/M_{Pl}\sim 10^{-17}$, however,  poses the much more difficult Standard Model (SM) \emph{naturalness} problem \cite{Wilson:1970ag,*Gildener:1976ai,*Weinberg:1978ym,Wells:2013tta}  coming from the fact that for scalar fields one would expect Planck scale corrections to the Higgs mass parameter (which is not protected by any symmetry) to force $v/M_{Pl}\sim 1$. 
Attempts to cure this UV sensitivity of the SM by modifying it at the weak scale lead us to expect either (a) a light Higgs in conjunction with new weakly coupled particles around the EW scale (supersymmetry) or (b) a composite Higgs emerging from strong dynamics. However, the experimental observation of only the SM Higgs with a somewhat intermediate mass and nothing else forces both ideas in rather uncomfortable corners of parameter (and theory) space. On the other hand (and quite intriguingly), if one extrapolates the SM up to the Planck scale, the experimentally measured value of the Higgs mass of $m_H=\unit[125.6]{GeV}$ is consistent with the nearly critical value of $\lambda(M_{Pl})\approx 0$ \cite{Holthausen:2012fk,Degrassi:2012ry,*Buttazzo:2013uya,Bezrukov:2012sa}, i.e. a vanishing Higgs self-interaction at the Planck scale, which could have interesting theoretical implications (see e.g. \cite{Shaposhnikov:2009pv,Holthausen:2012fk}). This observation warrants a reexamination of the hierarchy problem \cite{Bardeen:1995kv, Meissner:2006zh, *Meissner2008,Meissner:2009gs, *Holthausen:2009uc,Holthausen:2012fk,*Giudice:2013yca}. If there were a symmetry enforcing also $m_H(M_{Pl})\approx 0$, then since the renormalisation group running of the Higgs mass is multiplicative, 
$$
\frac{d\, m_H^2}{d\, \ln \mu}=\frac{3m_H^2}{8\pi^2}\left( 2\lambda +y_t^2-\frac{3g_2^2}{4}-\frac{3g_1^2}{20}\right),
$$
the smallness of the Higgs mass would be protected (i.e. it would stay small, if the Planck scale boundary condition would set it so), as long as there are no additional new physics scales between the Planck and EW scales\footnote{An additional scale would unavoidably reintroduce the need to fine-tune away large threshold corrections.}. Vanishing mass parameters may be motivated by \emph{classical scale invariance} of the particle physics action that emerges from Planck scale physics in some way which we will assume here\footnote{For related models using classical conformal symmetry, see \cite{Coleman:1973jx,*Gildener:1976ih,Sher:1988mj,Fatelo:1994qf,*Hempfling:1996ht,*Hambye:1995fr,*Foot:2007ay,*Foot:2007as,*Chang:2007ki,Meissner:2006zh,
*Meissner2008,*Foot:2007iy,*Hambye:2007vf,*Meissner:2009gs,*Iso:2009ss,
*Iso:2009nw,*Holthausen:2009uc,*Foot:2011xq,*Khoze:2013uia,*Kawamura:2013kua,*Gretsch:2013ooa, *Carone:2013wla, *Khoze:2013oga,*Iso:2012jn,*Englert:2013gz, *Farzinnia:2013pga,*Kawamura:2013kua, *Khoze:2013uia, *Foot:2013hna, Hur:2007uz,Hur:2011sv,Heikinheimo:2013fta}.}. This is a strong assumption which needs to be justified in a complete theory of quantum gravity and we refer the reader to some work in this 
direction in the literature \cite{Meissner2008,Meissner:2009gs}. Since it is not possible to reliably calculate Planck scale threshold corrections, the boundary condition $m_H(M_{Pl})= 0$ has to be taken as an assumption at this point. We here take the viewpoint \cite{Meissner2008,Meissner:2009gs} that the classical conformal symmetry value of $m_H(M_{Pl})= 0$ may be more easily justified in a complete theory than the usual Standard Model extrapolation of $m_H(M_{Pl})\sim 10^{-17} M_{Pl}\sim \unit[100]{GeV}$.
A solution of the hierarchy problem, as seen from the EW scale, would thus be directly connected to the conformal symmetry properties of a Planck scale embedding.
We assume conformal symmetry to act in such a way in the Planck scale UV completion of the theory that the Planck scale effectively does not enter as a physical scale into the particle physics action. In this case, of course, the old argument by Bardeen \cite{Bardeen:1995kv} can be applied, stating that the naive quadratic divergencies are spurious and only logarithmic terms related to the conformal anomaly survive.

The proposed scenario does, however, not work for the pure Standard Model due to the observed low energy parameters: The large top coupling makes Coleman-Weinberg symmetry breaking \cite{Coleman:1973jx} not possible \cite{Sher:1988mj, Lindner:1988ww} and new (bosonic) degrees of freedom have to be introduced to stabilize the potential. Even if the top mass were low enough then this would still not work, since Coleman-Weinberg symmetry breaking would lead to a Higgs mass which is too small. This implies that some new fields must be added in order to realize these ideas, i.e. it  unavoidably predicts new physics at accessible energy scales. Contrary to that there cannot be any intermediate scale physics coupling sizeably to the Higgs sector\footnote{Very weakly coupled models  such as low-to intermediate scale seesaw models do not give a large threshold correction to the Higgs mass  \cite{Casas:1999cd}. }.

If we accept the proposition of classical scale invariance of the particle physics action in conjunction with a direct Planck scale embedding, then there are a couple of aspects which might act as a guide to model building in this direction:
\begin{itemize}
\item The hierarchy between the QCD and EW scales is rather mild, for which reason it might be a good idea to have similar origin of both scales, namely the condensation in a  strongly coupled sector.
\item Since there is strong indication for dark matter (DM), and since if the DM scale close to EW scale, thermal freeze-out can produce right abundance of DM (the so-called WIMP miracle), it might be interesting to consider a scenario where both scales originate from a QCD-like condensation in a hidden sector. 
\end{itemize} 

We consider the dynamical details of a model proposed earlier in \cite{Hur:2007uz,Hur:2011sv,Heikinheimo:2013fta} which consists of a hidden $\SU{3}_{\rm H}$ gauge sector coupled via a real singlet scalar $S$ via a Higgs portal interaction to the SM:
\begin{align}
{\cal L}_{\rm H}
&=-\frac{1}{2}\mbox{Tr}~F^2+
\mbox{Tr}~\bar{\psi}(i\gamma^\mu D_\mu -
y S)\psi~,
\label{eq:LH}
\end{align}
where the hidden sector fermion $\psi$ transforms as a fundamental representation of $\SU{3}_{\rm H}$.
The trace is taken over the flavor as well as the color indices.
The ${\cal L}_{\mathrm{SM}+S}$ part of the total Lagrangian ${\cal L}_T ={\cal L}_{\rm H}+{\cal L}_{\mathrm{SM}+S} $ contains the 
SM gauge and Yukawa  interactions
along with the scalar potential
\be
V_{\mathrm{SM}+S}
&=&
\lambda_H ( H^\dag H)^2+
\frac{1}{4}\lambda_S S^4
-\frac{1}{2}\lambda_{HS}S^2(H^\dag H)~,
\label{eq:VSM}
\ee
where $H^T=( H^+ ~,~(h+iG)\sqrt{2}  )$ is the SM Higgs doublet field,
and $H^+$ and $G$ are the would-be Nambu-Goldstone fields.
Note that in our Lagrangian no mass term is present and 
all the coupling constants are dimensionless as required by classical scale invariance.
The classical scale invariance is quantum 
mechanically violated: It is broken not only by perturbative corrections as in the famous Coleman-Weinberg mechanism \cite{Coleman:1973jx} or equivalently by the non-vanishing $\beta$-functions, but also by the non-perturbative effect of dynamical chiral symmetry breaking.
It is this chiral symmetry breaking that generates a robust scale which is transferred into the SM sector through the singlet $S$, triggering the EW phase transition by generating the mass term for Higgs potential via the Higgs portal. We will exploit the similarity of this model to QCD to analyze non-perturbative properties such as confinement and chiral symmetry breaking. Furthermore we assume 3 flavors of hidden fermions whose chiral $\SU{3}_L\times \SU{3}_R$ symmetry is explicitly broken to the diagonal $\SU{3}_V$ by the hidden Yukawa coupling $y$. After chiral symmetry breaking, the dark pion pseudo-Nambu-Goldstone bosons of the model are naturally stable due to this unbroken symmetry and -depending on the model parameters- they might be viable cold DM candidates. No ad hoc discrete symmetry for the dark sector is needed.
 
 Similar models have been discussed previously in the literature, however, we go beyond these discussions in significant ways. Previous publications \cite{Hur:2007uz,Hur:2011sv,Heikinheimo:2013fta} have used linear and nonlinear sigma models for an effective description of the strongly interacting hidden sector at low energy.
We here use the Nambu-Jona-Lasinio (NJL) model \cite{Nambu:1961tp,*Nambu:1961fr}, which has the advantage of being able to dynamically describe the influence of the SM Higgs condensate on the dark sector dynamical condensate, and vice versa. The NJL model furthermore allows us to calculate the pion-pion-singlet coupling needed for the determination of the relic abundance produced via thermal freeze-out and allows a reliable calculation of the hidden chiral phase transition in the early universe. The scarcity of parameters in Eq.~(\ref{eq:LH}) in conjunction with NJL techniques allows us to predict the dark-matter nucleon cross-section as a function of the DM mass. Contrary to previous analyses, we also take seriously the requirement that the model should survive up to the Planck scale, which is a necessary condition for the assumed Planck scale 'solution' of the hierarchy problem. Combining this requirement with the NJL techniques and a assumed upscaled QCD, a sizable 
amount of the parameter space of the linear and nonlinear sigma 
models can be ruled out, as there the parameters are usually varied independently.
 
Let us briefly relate our work to alternative approaches which are similar in spirit: Hambye and Strumia \cite{Hambye:2013dgv} have recently discussed an $\SU{2}$ theory without fermions, spontaneously broken via Coleman-Weinberg, which has an automatically stable vector DM candidate (for a similar discussion, see also \cite{Heikinheimo:2013fta}). Bai and Schwaller \cite{Bai:2013xga} have discussed a model where the dark QCD scale is related to the visible QCD scale, however they have not discussed the EW scale. Buckley and Neil \cite{Buckley2013-gm} have discussed a hidden sector where the DM  candidate is baryon-like without assuming classical scale invariance.

The outline of the paper is as follows: We begin with a description of the model and NJL formalism in chapter \ref{sec:model}, discuss DM phenomenology in chapter \ref{sec:dark-matter}, briefly describe the nature of the phase transition in chapter \ref{sec:phase transition} and conclude in chapter \ref{sec:summary}.

\section{The model and its Effective Lagrangian}\label{sec:model}
\subsection{NJL treatment of the low-energy theory}
To treat the dynamical chiral symmetry breaking,  
we will use a particular effective description, namely the Nambu-Jona-Lasinio (NJL) model
 \cite{Nambu:1961fr}. In analogy with QCD, we can use as a low-energy approximation of (\ref{eq:LH}) the NJL Lagrangian
\be
{\cal L}_{\rm NJL}&=&\mbox{Tr}~\bar{\psi}(i\gamma^\mu\partial_\mu -
y S)\psi+2G~\mbox{Tr} ~\Phi^\dag \Phi
+G_D~(\det \Phi+h.c.)~,
\label{eq:NJL10}
\ee
where 
\be
\Phi_{ij}&=& \bar{\psi}_i(1-\gamma_5)\psi_j=
\frac{1}{2}
\lambda_{ji}^a \mbox{Tr}~\bar{\psi}\lambda^a(1-\gamma_5)\psi\nn\\
(\Phi^\dag)_{ij}&=&
 \bar{\psi}_i(1+\gamma_5)\psi_j=
\frac{1}{2} \lambda_{ji}^a \mbox{Tr}~\bar{\psi}\lambda^a(1+\gamma_5)\psi~,
\ee
and $\lambda^a$ are the Gell-Mann matrices with
$\lambda^0=\sqrt{2/3}~{\bf 1}$.
The last term in (\ref{eq:NJL10}) is present due to chiral anomaly
of the axial $U(1)_A$ (or instanton effect) \cite{Kobayashi:1970ji,*Kobayashi:1971qz,*Hooft:1976up}, and
it breaks $U(1)_A$ down to $Z_3$ (for $n_f=3$),
implying that the Lagrangian ${\cal L}_{\rm NJL}$ 
has a global symmetry
$SU(3)_V\times U(1)_V\times Z_3$. As noted earlier, the  chiral symmetry $SU(3)_L\times SU(3)_R$
is explicitly broken by the Yukawa coupling with the singlet $S$.
The effective Lagrangian ${\cal L}_{\rm NJL}$ has four parameters;
$y~,~G~,~G_D$ and the cutoff $\Lambda$
\footnote{We need a cutoff $\Lambda$  because
${\cal L}_{\rm NJL}$ contains non-normalizable
interactions.}, which have canonical dimensions of 
$0,-2$, $-5$, and $1$ respectively.
Since the original Lagrangian ${\cal L}_{H}$ has only two 
independent parameters, the parameters $G~,~G_D$ and  $\Lambda$
are not independent and can be related by the NJL approach. We will use relations from observed hadron physics which we will then scale up to obtain the NJL parameters.

To deal with the non-renormalizable Lagrangian (\ref{eq:NJL10}) we will use a  self-consistent mean-field (SCMF) approximation which has been intensely studied by Hatsuda and Kunihiro in the past
\cite{Kunihiro:1983ej,Kunihiro:1987bb,Hatsuda:1994pi}. While the general features of the model in Eq.~\eqref{eq:LH} should be similar for any number of dark color $n_c$ and hidden flavor $n_f$ (as long as the theory is asymptotically free and confining), we here restrict ourselves to $n_f=n_c=3$, which allows us to make the rough but justifiable estimation that we can approximately use (up to an overall scale) the values of 
$G~,~G_D$ and  $\Lambda$ that correspond to the  real world
of hadrons.  (In contrast to \cite{Kunihiro:1983ej,Kunihiro:1987bb,Hatsuda:1994pi},
we use a four-dimensional cutoff $\Lambda$. )
This allows us to eliminate the extra free parameters. 
Under this assumption, we calculate the actual values for
$G~,~G_D$ and  $\Lambda$  in the Appendix \ref{sec:appendix}. Here we briefly outline this approximation method
\cite{Kunihiro:1983ej,Kunihiro:1987bb,Hatsuda:1994pi}.

One assumes that the dynamics of the theory creates a chiral symmetry breaking condensate 
\be
\langle0 \vert \bar{\psi}_i \psi_j \vert 0 \rangle \equiv\widehat{\bar{\psi}_i \psi_j} &=&
-\frac{1}{4G}  \diag(\sigma,\sigma,\sigma)~,
\label{eq:bbb}
\ee
which is treated as a classical field $\sigma(x)$. Since we assume the explicit breaking of the $\SU{3}_L\times \SU{3}_R$ flavor symmetry to be small, the other important effective fields are given by the dark pions
\be
\phi_a &=&
 -2i G~\widehat{\bar{\psi}\gamma_5\lambda^a \psi}~.
 \label{eq:phi1}
\ee
We thus restrict our discussion (in a more complete treatment, one may add terms involving $\eta$ or $\rho$ mesons) to the mean fields collected in 
 \be
 \widehat{\Phi} &=&\varphi=
-\frac{1}{4G}  \left(\diag(\sigma,\sigma,\sigma)+i(\lambda^a)^T \phi_a\right)~.
\label{eq:varphi}
 \ee
In the self-consistent mean field approximation one splits up the NJL Lagrangian (\ref{eq:NJL10}) into the sum
$$
\mathcal{L}_{\rm NJL}= \mathcal{L}_{0}+\mathcal{L}_{I}~,
$$
where $\mathcal{L}_{0}$ describes the mean field dynamics  and $\mathcal{L}_{I}$ describes the rest (i.e. the interactions that form the condensate etc.). The self-consistency requirement forces this part of the Lagrangian to vanish in the assumed vacuum:
$$
\langle 0\vert \mathcal{L}_{I}\vert 0\rangle =0.
$$
After some manipulations, which we relegate to the Appendix \ref{sec:appendix}, one finds the following form for ${\cal L}_0$:
\be
{\cal L}_0 & =&
i~\mbox{Tr}\bar{\psi}\gamma^\mu\partial_\mu\psi-
\left(\sigma+y S-\frac{G_D}{8G^2}\sigma^2\right)
\mbox{Tr}\bar{\psi}\psi -i \mbox{Tr}\bar{\psi}\gamma_5 \phi\psi -\frac{1}{8G}\left(3\sigma^2
+2\sum_{a=1}^8\phi_a\phi_a\right)\nn\\
& &+\frac{G_D}{8G^2}\left(
- \mbox{Tr}\bar{\psi}\phi^2 \psi+\sum_{a=1}^8\phi_a\phi_a \mbox{Tr}\bar{\psi}\psi +i \sigma\mbox{Tr}\bar{\psi}\gamma_5 \phi \psi+ \frac{\sigma^3}{2G}+
\frac{\sigma}{2G} \sum_{a=1}^8(\phi_a)^2\right).
\label{eq:L0}
\ee
This Lagrangian determines the dynamics of the effective condensate fields and we will use it to calculate the effective potential
at zero and finite temperature, the
DM mass $m_{\rm DM}$ and the 
DM-DM-singlet $S$ coupling, which determines the DM relic abundance.

Note that integrating out the fermion fields at the one-loop order
produces corrections of $O(n_c)$.
If we rescale $G$ as $G\to G/n_c$ and $G_D$ as $G_D\to G_D/n_c^2$,
we find these one-loop order corrections are indeed  the leading order
corrections in $1/n_c$ expansion.

\subsection{The Effective Potential, Symmetry Breaking and Scalar Masses}
In the model we are considering here, there are in principle two ways in which the quantum level breaking of classical scale invariance may manifest itself. Through the RG evolution, the scalar potential may develop a flat direction and quantum corrections then shift the scalar VEV to a non-vanishing value a la Coleman-Weinberg \cite{Coleman:1973jx,Gildener:1976ih}. The other possibility is the one we are focusing on here, namely the case that the additional gauge interaction grows strong and dynamically sets a condensation scale, as happens for QCD. 

To be certain that symmetry breaking proceeds in this (and not in the Coleman- Weinberg) way, one has to study the RG evolution of the scalar potential parameters and make sure that the stability conditions
$$
4\lambda_H \lambda_S-\lambda_{HS}^2>0, \qquad \lambda_H>0, \qquad \lambda_S>0
$$
are fulfilled\footnote{Actually, the stability conditions have to larger than a typical 1-loop contribution, e.g. a weak gauge coupling to the power of four, in the case of the Higgs field \cite{Gildener:1976ih}.} until the confinement scale, where the coupling of strong hidden sector $g_4$ grows large. This situation, where the gauge instability (confinement) happens before the vacuum potential instability (Coleman-Weinberg), is  realized in a wide range of parameters, as we will discuss later.
\begin{figure}
\begin{center}
\includegraphics[width=10cm]{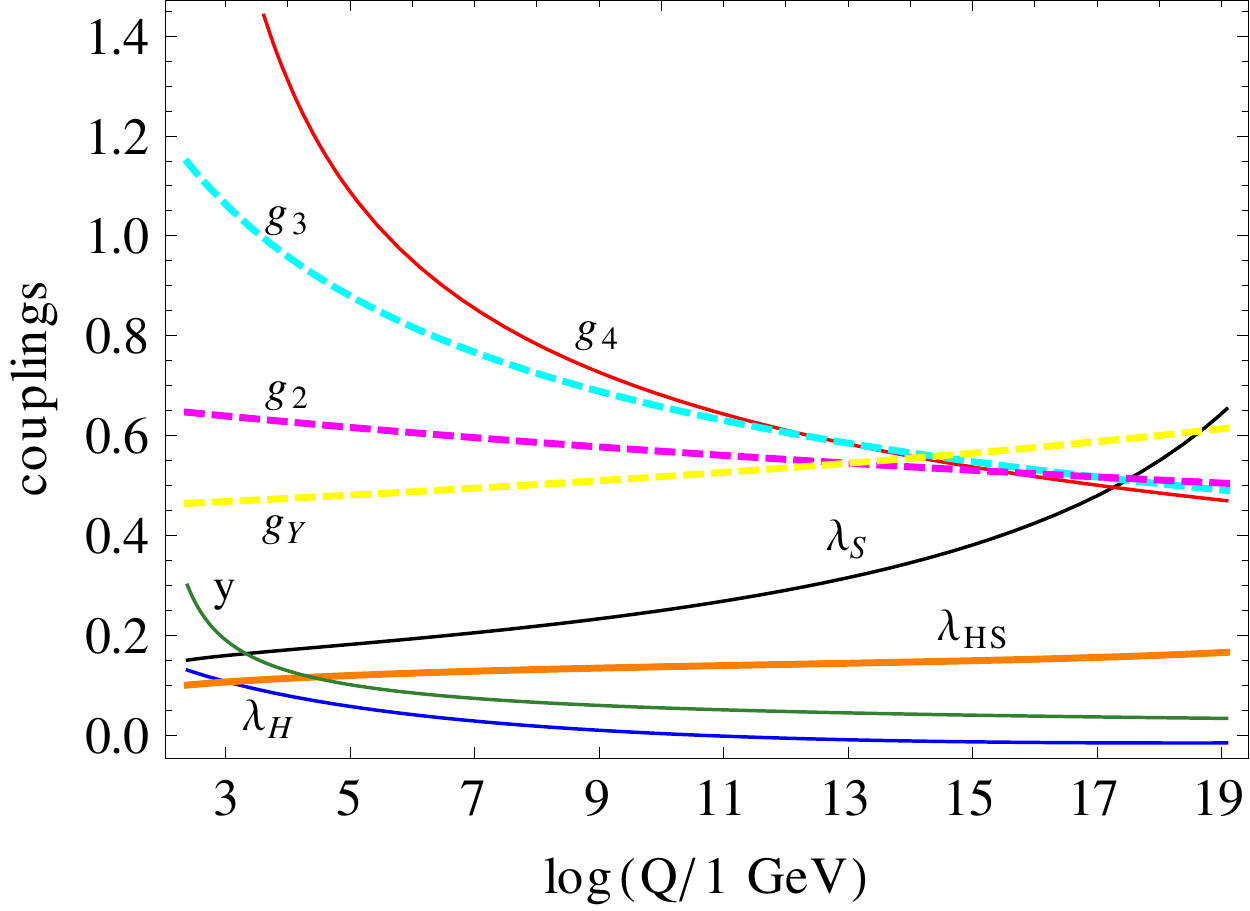}
\end{center} 
\caption{
The RG evolution of an exemplary set of values of model parameters for different energy scale are shown with the boundary values at $Q=\unit[1]{TeV}$ given as $g_4^2=4\pi$, $\lambda_H=0.13$, $\lambda_S=0.15$, $\lambda_{HS}=0.1$, $y=0.3$. The SM gauge coupling constants are denoted as $g_Y=\sqrt{5/3}g_1$, $g_2$ and $g_3$.}
\label{fig:running-sm}
\end{figure}
In Fig.~\ref{fig:running-sm} we show the running of the relevant couplings in our model framework. The SM gauge coupling constants are denoted as $g_Y=\sqrt{5/3}g_1$, $g_2$ and $g_3$. The rest of the couplings are set at $Q=\unit[1]{TeV}$ and the values are given as $g_4^2=4\pi$, $\lambda_H=0.13$, $\lambda_S=0.15$, $\lambda_{HS}=0.1$, $y=0.3$. The gauge coupling $g_4$ possesses a similar value to QCD gauge coupling $g_3$ at the Planck scale. From aesthetic point of view this observation is intriguing, as this might provide a strong support to our argument that the hierarchy of the QCD scale and strong hidden sector scale (EW scale) is mild due to the common origin in Planck scale. As one goes to smaller energies, the strong hidden sector coupling grows non-perturbative at a higher scale than QCD, due to the smaller number of flavors. From the dark matter perspective this observation is also fascinating, as the similarity of both the strong sectors could explain why the relic abundance of DM is around the 
same 
order of magnitude in comparison to the abundance of baryons. 

The dominant mechanism of dimensional transmutation is therefore the condensation in the hidden sector.
In the NJL picture, the condensation can be studied using the one-loop effective potential, which can be obtained by
integrating out the fermion fields in 
${\cal L}_0$ given in Eq.~(\ref{eq:L0}):
\be
V_{\rm NJL}(\sigma,S)
&  &= \frac{3}{8G}\sigma^2-
\frac{G_D}{16G^3}\sigma^3-3 n_c I_0(M,0)~,
\label{eq:Vnjl}
\ee
where $I_0(M,p^2)$ is given in Eq.~(\ref{eq:I0}) and
the ``constituent mass'' $M$ is given by
\be
M& =&\sigma+yS-\frac{G_D}{8G^2}\sigma^2~.
\label{eq:cmass0}
\ee
The integral $I_0$ is evaluated with a four-dimensional momentum cutoff as the NJL framework is an effective field theory. In fact all the loop integrals that we will encounter later are computed with four-dimensional momentum cutoff $\Lambda$.
The potential is asymmetric in $\sigma$, which is a consequence of the anomaly term (the last term)
in the NJL Lagrangian (\ref{eq:NJL10}).
We will be mostly concerned with the regime of small $y$, where the back reaction of the $S$ condensate onto the chiral condensate $\sigma$ may be neglected. We assume the parameters $G$, $G_D$ and the phenomenological cutoff $\Lambda$ to be rescaled from their QCD values, i.e.
$(2G^{\mathrm{QCD}})^{-1/2}=326$ MeV,
$(-G_D^{\mathrm{QCD}})^{-1/5}=437$ MeV  and
$\Lambda^{\mathrm{QCD}}=924$ MeV
(which we have calculated for real-world QCD in the Appendix \ref{sec:appendix}) according  to their dimensions as
\be
G= f^{-2} G^{\mathrm{QCD}}~,~G_D=  f^{-5} G_D^{\mathrm{QCD}}~,~\Lambda= f\Lambda^{\mathrm{QCD}}~
\label{eq:fdef}
\ee
by a common rescaling factor $f$, which is determined from the requirement that $ \langle h\rangle=\unit[246]{GeV} $. As an example we obtain the minimum of $V_{\mathrm{SM}+S}+V_{\rm NJL}$ of QCD, i.e.
$\langle h\rangle^{\rm QCD} =\unit[0.021]{GeV}$, $\langle S\rangle^{\rm QCD} =\unit[0.107]{GeV}$, and  
$\langle \sigma\rangle^{\rm QCD} =\unit[0.280]{GeV}$
with the following parameters
\begin{align}
y&=0.0052~,
\lambda_{H}=0.13~,
~\lambda_{HS}=0.01~,~\lambda_{S}=0.19~,
\label{eq:exm1}
\end{align}
where we can determine $f=\unit[246]{GeV}/\langle h \rangle^{\rm QCD} \approx 11760$ to scale up all relevant parameters used in the strongly coupled sector and the singlet.

In the case of a small $y$ we can neglect back-reactions on $\sigma$ in Eq.~(\ref{eq:cmass0}) and find 
\be
\langle \sigma \rangle =f \times \langle \sigma \rangle ^{\mathrm{QCD}}=f \times 0.280 \GeV~.
\ee
In the same limit we can treat the Yukawa coupling as an external source for $V_{\mathrm{SM}+S}$ of Eq.~(\ref{eq:VSM}) and consider 
\be 
V_{\rm SP}=V_{\mathrm{SM}+S}-y \frac{3}{4 G} \langle \sigma \rangle S, 
\ee
from which follows
\begin{align}
\frac{\langle h\rangle ^2}{\langle S\rangle ^2}&=\frac{\lambda_{HS}}{2 \lambda_H}, \nn\\
\langle h \rangle ^2 = \frac{\lambda_{HS}}{2 \lambda_H} \left[ \frac{3 y \sigma \lambda_H}{G (4 \lambda_H \lambda_S- \lambda_{HS}^2)} \right]^{2/3}&=f^2 \times \frac{\lambda_{HS}}{2 \lambda_H} \left[ \frac{3 y \sigma^{\mathrm{QCD}} \lambda_H}{G^{\mathrm{QCD}} (4 \lambda_H \lambda_S- \lambda_{HS}^2)} \right]^{2/3}.
\end{align}
Obviously, small values for $\lambda_{HS}$ and $y$ imply a large hierarchy between the various scales, which allows to determine the rescaling factor $f$ from the requirement that $ \langle h\rangle=\unit[246]{GeV} $.  In the case where the backreaction cannot be neglected anymore, one has to consider the full coupled potential, which can only be done numerically. Throughout this paper we consider only the full coupled potential $V_{\mathrm{SM}+S}+V_{\rm NJL}$ and compute relevant quantities numerically. 

The next step is to obtain the mass spectrum of particles in the model.
In our system we have $h, S, \sigma$ as CP-even scalars, while the DM is CP-odd.
The CP-even scalars, $h, S, \sigma$ , mix with each other  and the SM Higgs-like particle found in ATLAS  \cite{Aad:2012tfa} and CMS \cite{Chatrchyan:2012ufa}  has to be identified with one of the mass eigenstates.

Once the absolute minimum is determined, we are all set to calculate the scalar mass spectrum.
Note that $h$ and $S$ are propagating fields at the tree-level,
but $\sigma$, as well as the DM field $\phi_a$, becomes a dynamical field in the one-loop
order ($\mathcal{O}(n_c)$). Therefore, $\sigma$ does not have a canonically
normalized kinetic term even in the lowest order.
We therefore have to consider inverse propagators
$\Gamma_{ij}~(i,j=h,S,\sigma)$.
At $\mathcal{O}(n_c)$ there are contributions to
$\Gamma_{SS}, \Gamma_{S\sigma}$ and $\Gamma_{\sigma\sigma}$:
\begin{align}
\Gamma_{hh}(p^2) =& p^2-3\lambda_{H}\langle h\rangle^2+\frac{1}{2}\lambda_{HS}\langle S\rangle^2,\quad\Gamma_{hS} =\lambda_{HS}\langle h\rangle\langle s\rangle,\quad\Gamma_{h\sigma} =0,\nn \\
\Gamma_{SS} (p^2)=&p^2-3\lambda_{S}\langle S\rangle^2+\frac{1}{2}\lambda_{HS}\langle h\rangle^2-3 n_c y^2 I_4 (p^2,\langle M\rangle),\nn\\
\Gamma_{S\sigma}(p^2) =&-3 n_c y(1-G_D \langle\sigma \rangle /4 G^2)I_4 (p^2,\langle M\rangle),\nn\\
\Gamma_{\sigma\sigma}(p^2) =&-\frac{3}{4G} +\frac{3 G_D \langle \sigma\rangle}{8G^3} -3 n_c \left(1-G_D \langle\sigma \rangle/4 G^2\right)^2 I_4 (p^2,\langle M\rangle)+\nn\\
& +3 n_c \frac{G_D }{G^2}I_2(\langle M\rangle),
\label{eq:gammas}
\end{align}
where the function $I_2 (M)$ is defined in (\ref{I2}), and 
\be
I_4(p^2,M)&= &
\int \frac{d^4 k}{i(2\pi)^4}
\frac{\mbox{Tr}(k+M)
(k\hspace{-0.2cm}/-p\hspace{-0.2cm}/+M)}
{(k^2-M^2)((k-p)^2-M^2)}.
\ee
The propagator matrix $\Delta_{ij}(p^2)=i(\Gamma^{-1})_{ij}(p^2)$ has to be diagonalized and the physical mass spectrum can be obtained from the pole of such diagonalized propagators.
Once the poles $\tilde{m}_1^2,~\tilde{m}_2^2$ and
$\tilde{m}_3^2$ are found, we can compute the corresponding  eigenvectors
$\xi^{(i)}$ from
\be
\Gamma_{ij}(\tilde{m}_k^2)~\xi^{(k)}_j&=&0~.
\ee
For the parameters given in (\ref{eq:exm1}), $y=0.0052$ along with
the corresponding rescaling for $G, G_D$ and $\Lambda$,
we find 
\begin{align}
\tilde{m}_1 &= m_h=\unit[125.4]{GeV}, &(\xi^{(1)})^T&=(0.999,\;0.004,\;3\times10^{-5}),\nn\\
\tilde{m}_2 &= m_S= \unit[946.4]{GeV}, &(\xi^{(2)})^T&=(-0.004,\;0.999,\;0.008),\nn \\
\tilde{m}_3 &= m_\sigma= \unit[6833]{GeV}, &(\xi^{(3)})^T&=(0,\;-0.0025,\;1.000).
\end{align}
The flavor eigenstates $(h,S,\sigma)$ and the mass 
eigenstates $(s_1,s_2,s_3)$ are
related by
\be
\left(\begin{array}{c}
h\\S\\ \sigma
\end{array}\right) &=&\left(\begin{array}{ccc}
\xi^{(1)}_1 & \xi^{(2)}_1 & \xi^{(3)}_1\\
\xi^{(1)}_2 & \xi^{(2)}_2 & \xi^{(3)}_2\\
\xi^{(1)}_3 & \xi^{(2)}_3 & \xi^{(3)}_3
\end{array}\right)\left(\begin{array}{c}
s_1\\s_2\\ s_3
\end{array}\right)~.
\label{eigenstate}
\ee
From the example parameters chosen above we obtain the Higgs mass value close to the experimentally measured value. The next task is to find the parameter space for $\lambda_H$, $\lambda_S$, $\lambda_{HS}$ and $y$ which predict a set of experimental observables that are still allowed by collider experiments and dark matter searches.

\subsection{Bounds from requiring survival up to the Planck scale}

Before we perform a scan of parameters, the parameter space can be constrained by the following assumptions: As we assume that the SM with the hidden sector is scale invariant up to the Planck scale, all parameters have to be perturbative up to the Planck scale in accordance to the renormalization group equations. This crucial assumption constrains the allowed parameter region of $\lambda_H$, $\lambda_S$, $\lambda_{HS}$ and $y$. The one-loop beta functions for the hidden sector and modified SM are given as 
\begin{align}
&16\pi^2 \beta_{\lambda_H}=\lambda_H(-9g_2^2-3g_1^2+12y_t^2)+24\lambda_H^2+\frac{3}{4}g_2^4+\frac{3}{8}(g_1^2+g_2^2)^2-6y_t^4+\frac{1}{2}\lambda_{HS}^2, \nn \\
&16\pi^2 \beta_{\lambda_{HS}}=-2\lambda_{HS}\left(2\lambda_{HS}-3\lambda_S+\frac{9}{4}g_2^2+\frac{3}{4}g_1^2-3y_t^2-6\lambda_H-18y^2 \right),\nn\\
&16\pi^2 \beta_{\lambda_S}=2\lambda_{HS}^2+18\lambda_S^2+72y^2\lambda_S-18y^4,\nn\\
&16\pi^2 \beta_y=3y(7y^2-4g_4^2),\nn\\
&16\pi^2 \beta_{g_4}=-9g_4^3,
\end{align}
with the rest of the SM RGE remained unchanged. 

\begin{figure}[t]
\begin{center}
\includegraphics[width=10cm]{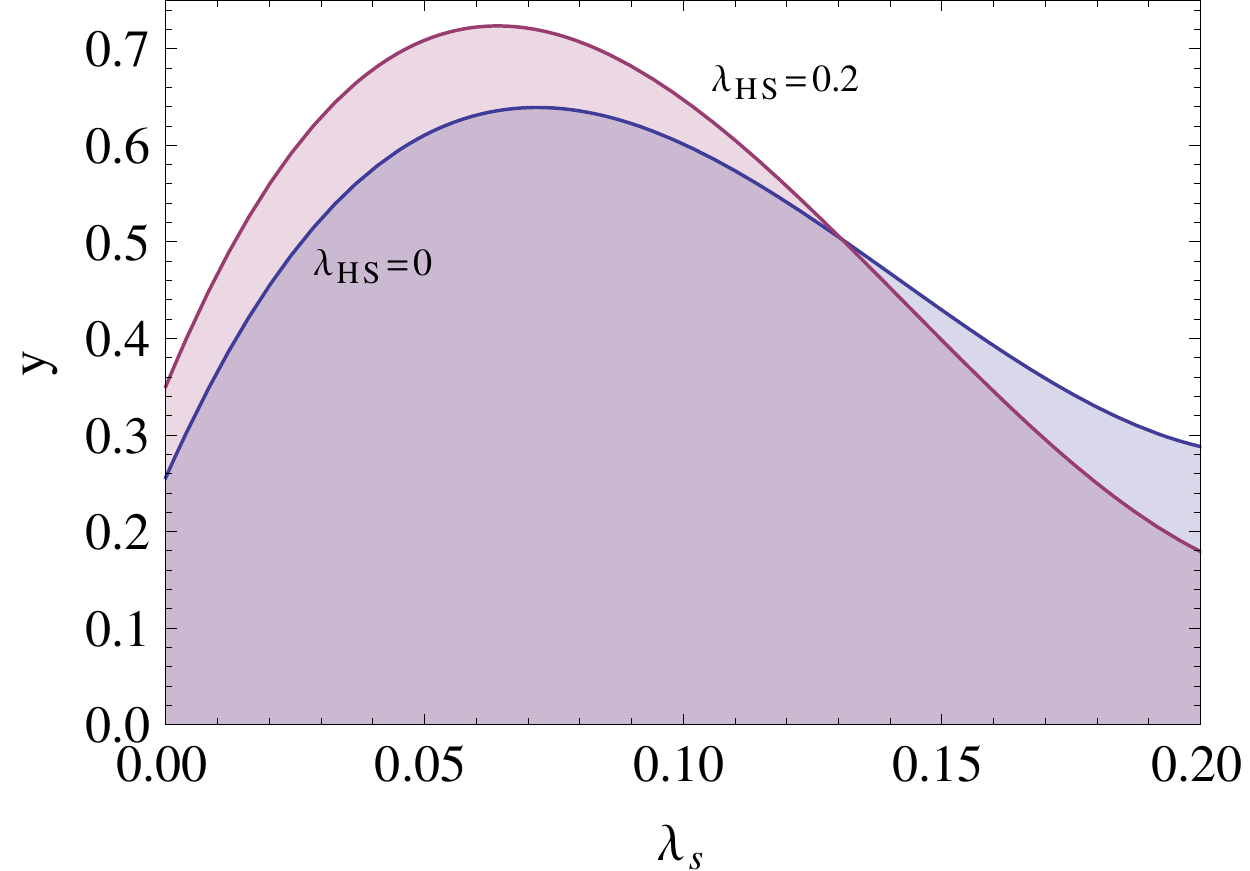}
\end{center} 
\caption{The allowed regions for the parameters $\lambda_S$ and $y$ (the $Q=\unit[1]{TeV}$ scale) are shaded for different value of $\lambda_{HS}$.}
\label{fig:running}
\end{figure}

We can impose some of the boundary conditions of the hidden sector couplings based on theoretical reasoning. The hidden gauge sector is strongly interacting at the vicinity of $Q \approx \unit[1]{TeV}$, i.e. $g_4^2(Q)\approx4\pi$. The Higgs quartic coupling $\lambda_H$ can be obtained from the Higgs mass measurement \cite{Aad:2012tfa,Chatrchyan:2012ufa}. Although in the model the measured Higgs mass depends mainly on two parameters, $\lambda_H$ and $\lambda_{HS}$, lowering $\lambda_H(Q)<0.13$ will destabilize the Higgs potential while increasing $\lambda_H(Q)> 0.14$ will require a larger mixing with the $S$ field, which is strongly constrained. Therefore we have chosen $\lambda_H(Q)\approx 0.13$ for the rest of our analysis.

The rest of the couplings, i.e. $\lambda_{HS}$, $\lambda_S$ and $y$ have to be determined from the RGE, without the couplings hitting a Landau pole or destabilizing the potential. At one-loop order the beta function of the Yukawa coupling $y$ only receives contributions from $y$ and $g_4$. With the boundary condition of $g_4$ imposed, the range of $y$ valid up to Planck scale is naively determined to be $y(Q)\in \left(0,0.6 \right)$. However as $y$ also contributes to the running of $\lambda_S$, its range is strongly constrained by the perturbativity and vacuum stability 
of $S$. We found that $\lambda_S(Q)\in \left(0, 0.2\right)$ are sufficient to guarantee the running up to the Planck scale, barring the two-loop beta function and threshold effect contributions. Refer to Fig.~\ref{fig:running} for a more accurate region of parameter space. Once the range of $\lambda_S$ is known, it is easy to determine the range of $\lambda_{HS}$ from the vacuum stability condition
\begin{align}
4\lambda_H \lambda_S-\lambda_{HS}^2>0 
\end{align}
which yields $\lambda_{HS}(Q)\in \left(0,0.2 \right)$. 

Once the parameter space is fixed, we can now calculate the masses and couplings for the scalars like the previous section and most importantly, we can determine the properties of our dark matter candidate.

\section{Properties of Dark Pions}\label{sec:dark-matter}
\subsection{DM mass and couplings}

\begin{figure}
  \includegraphics[width=9cm]{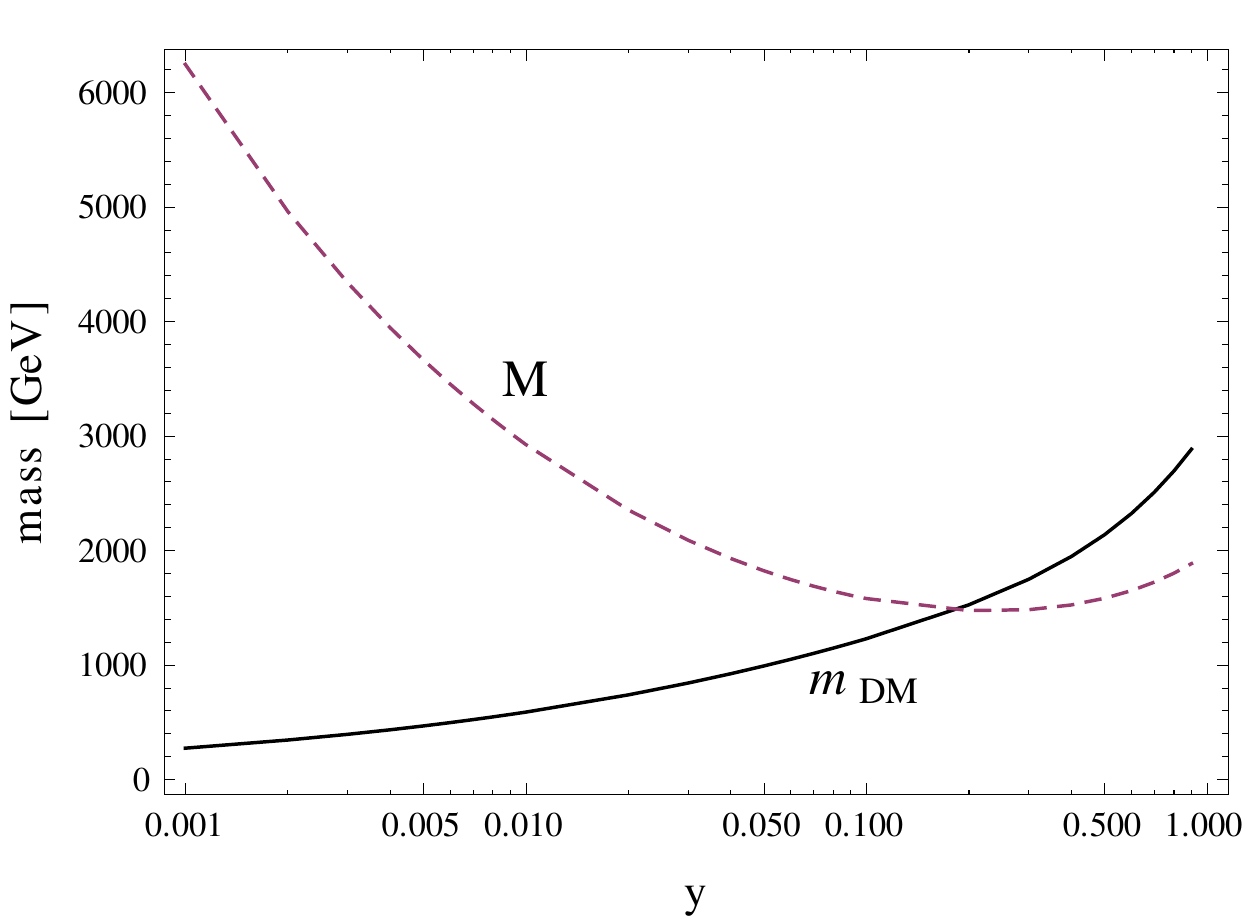}
\caption{\label{y-mDM}\footnotesize
The mass of DM $m_{\rm DM}$ and constituent quark mass $M$
as a function of $y$, where the scalar couplings are fixed
at the values given in Eq.~(\ref{eq:exm1}).}
\end{figure}

As we have mentioned above, our DM candidates are CP-odd scalars, i.e. the dark pions. We want to recall also that hidden sector baryons could be stable due to hidden baryon number conservation, therefore contributing to additional DM abundance. In our analysis we will ignore the hidden baryons, focusing on only the dark pions which are stable due to hidden sector flavor symmetry. Like the $\sigma$ field, the DM field $\phi_a$ has no tree level kinetic terms, its mass is generated at one-loop and it is defined as the zero of the inverse propagator:
\be
\Gamma_{\rm DM}(p^2)
&=&-\frac{1}{2G}+\frac{G_D}{8G^3}\langle \sigma\rangle
-\left(1-\frac{G_D}{8G^2}\langle \sigma\rangle
\right)^2~2 n_c I_1(p^2,\langle M\rangle )
+\frac{G_D}{G^2}n_c I_2(\langle M\rangle )~,
\label{sigma-dm}
\ee
where $I_1$ and $I_2$ are given in Eq.~(\ref{I2}), respectively, and the term $\langle M\rangle=\langle\sigma\rangle+y\langle S\rangle-G_D/8G^2\langle\sigma\rangle^2$
is given in Eq.~(\ref{eq:cmass0}). From the inverse propagator above
we can calculate the dark matter mass $m_{\rm DM}$ and the wave function renormalization constant $Z_{\rm DM}$:
\be
\Gamma_{\rm DM}(m_{\rm DM}^2)&=&0~,~
Z_{\rm DM}^{-1} = \frac{d \Gamma_{\rm DM}(p^2)}{d p^2}~\left|_{p^2=m_{\rm DM}^2}  \right.~.
\label{wave0}
\ee

\begin{figure}
  \includegraphics[width=6cm]{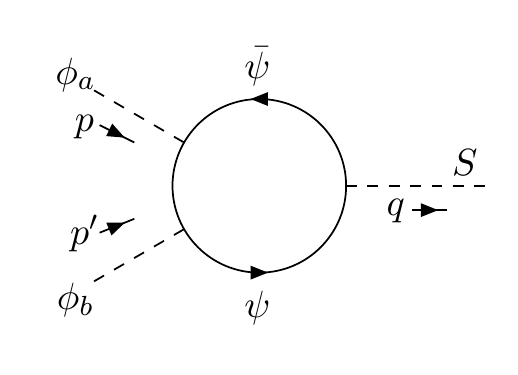}
    \includegraphics[width=5.7cm]{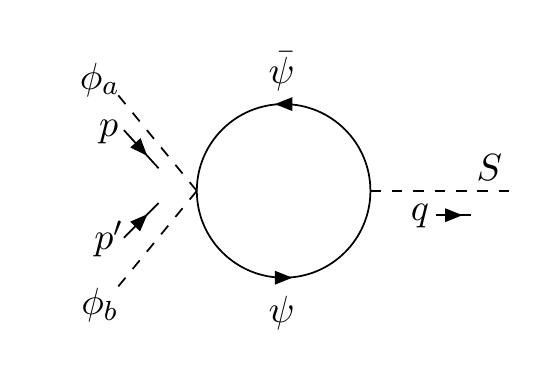}
\caption{\label{dmdms}\footnotesize
One-loop diagrams contributing
to the DM-DM-$S$ coupling.}
\end{figure}
The dark matter mass $m_{\rm DM}$ vanishes 
if $y=0$, due to the chiral symmetry that emerges in this limit.
For the minimum given in Eq.~(\ref{eq:exm1}) and $y=0.0052$, we obtain $m_{\rm DM}=\unit[473]{GeV}$, where the rescaling factor 
(defined in Eq.~(\ref{eq:fdef})) in this example is $f\simeq
11760$.
Fig.~\ref{y-mDM}
shows the DM mass $ m_{\rm DM}$ and constituent quark mass $M$
as a function of $y$, where the scalar couplings are fixed
at the values given in Eq.~(\ref{eq:exm1}). Note that the NJL approximation is only valid when $m_{\rm DM}<M$, as when the constituent mass $M$ is lighter we cannot integrate out the fermions. This observation will constrain our parameter space for $y$ later.

Before we calculate the annihilation cross section of our DM, we need to know how it communicates with the SM sector. It turns out that the dark pion is connected to the SM sector via the messenger scalar $S$ only through loop-suppressed interactions:
The DM-DM-$S$ coupling is generated from the one-loop diagrams shown in Fig.~\ref{dmdms}. We find that the three-point vertex function is given by
\be
\Gamma_{DM-DM-S}(p,p',M) &=&
2n_c y\left(1-\frac{G_D \langle\sigma\rangle}{8G^2}\right)^2
I_{5a}(p,p',M)+n_c y\frac{G_D }{4G^2}
I_{5b}(p,p',M)~,
\label{dm-dm-s}
\ee
where
\be
I_{5a}(p,p',M)&= &
\int \frac{d^4 k}{i(2\pi)^4}
\frac{\mbox{Tr}
(k\hspace{-0.2cm}/+M)
\gamma_5(k\hspace{-0.2cm}/-p\hspace{-0.2cm}/+M)
(k\hspace{-0.2cm}/+p'\hspace{-0.3cm}/+M)
\gamma_5}{((k-p)^2-M^2)(k^2-M^2)((k+p')^2-M^2)}~,\nn\\
I_{5b}(p,p',M)&= &
\int \frac{d^4 k}{i(2\pi)^4}
\frac{\mbox{Tr}(k\hspace{-0.2cm}/-p'\hspace{-0.3cm}/+M)
(k\hspace{-0.2cm}/+p\hspace{-0.2cm}/+M)}
{((k-p')^2-M^2)((k+p)^2-M^2)}~.
\ee
When computing the relic abundance of DM and its
cross section with matter, we will need
$\Gamma_{DM-DM-S}(p,p',M)$ for  $p=p'=(m_{\rm DM},{\bf 0})$
and for $p=-p'$, which are denoted by $\kappa_{s}$ and 
$\kappa_{t}$. (The integrals can be computed analytically for
these momentum configurations.) Using the expressions
\begin{align}
&\Gamma_{a}^s =\left.I_{5a}(p,p',M)\right|_{
p=p'=(m_{\rm DM},{\bf 0})}, &\Gamma_{a}^t=
\left.I_{5a}(p,p',M)\right|_{p=-p',p^2=m_{\rm DM}^2}, \nn \\
&\Gamma_{b}^s =\left.I_{5b}(p,p',M)\right|_{
p=p'=(m_{\rm DM},{\bf 0})}, &\Gamma_{b}^t=
\left.I_{5b}(p,p',M)\right|_{p=-p',p^2=m_{\rm DM}^2}~,
\end{align}
we obtain the couplings
\be
\kappa_s &=&
2n_c y\left(1-\frac{G_D \langle\sigma\rangle}{8G^2}\right)^2
\Gamma_{a}^s+n_c y\frac{G_D }{4G^2}
\Gamma_{b}^s~\label{kappas}~,\nn\\
\kappa_t &=&
2n_c y\left(1-\frac{G_D \langle\sigma\rangle}{8G^2}\right)^2
\Gamma_{a}^t+n_c y\frac{G_D }{4G^2}
\Gamma_{b}^t~.
\label{kappat}
\ee

\begin{figure}[t]
\begin{center}
\subfloat{
\begin{fmffile}{4point1}
	        \begin{fmfgraph*}(100,60)
		    \fmfcmd{vardef middir (expr p,ang) = dir(angle direction length(p)/2 of p +ang) enddef;
		      style_def arrow_right expr p = shrink(.7); cfill(arrow p shifted (4thick*middir(p,-90))); endshrink enddef;}  
	            \fmfleft{g1,g2}
		    \fmflabel{$\phi_b$}{g2}
                    \fmflabel{$\phi_a$}{g1}
	            \fmfright{h1,z1}
                    \fmflabel{$S$}{h1}
		    \fmflabel{$S$}{z1}
	            \fmf{dashes,label=$p'$,l.side=left}{g1,t1}
	            \fmf{dashes,label=$p$,l.side=left}{g2,t2}
		    \fmf{dashes,label=$q$,l.side=right}{h1,t3}
		    \fmf{dashes,label=$q'$,l.side=right}{z1,t4}
		    \fmf{plain,left=0.3}{t1,t2}
		    \fmf{fermion,right=0.5,label=$\psi$,l.side=right}{t1,t3}
		    \fmf{plain,right=0.3}{t3,t4}
		    \fmf{fermion,right=0.5,label=$\bar{\psi}$,l.side=right}{t4,t2}
		    \fmf{arrow_right}{g1,t1}
		    \fmf{arrow_right}{g2,t2}
		    \fmf{arrow_right}{t3,h1}
		    \fmf{arrow_right}{t4,z1}
	        \end{fmfgraph*}
	    \end{fmffile}}
\hspace{2cm}
\subfloat{
\begin{fmffile}{4point2}
	        \begin{fmfgraph*}(100,60)
		    \fmfcmd{vardef middir (expr p,ang) = dir(angle direction length(p)/2 of p +ang) enddef;
		      style_def arrow_right expr p = shrink(.7); cfill(arrow p shifted (4thick*middir(p,-90))); endshrink enddef;
		      style_def arrow_left expr p = shrink(.7); cfill(arrow p shifted (4thick*middir(p,90))); endshrink enddef;}
	            \fmfleft{g1,g2}
		    \fmflabel{$\phi_b$}{g2}
                    \fmflabel{$\phi_a$}{g1}
	            \fmfright{h1,z1}
                    \fmflabel{$S$}{h1}
		    \fmflabel{$S$}{z1}
	            \fmf{dashes,label=$p'$,l.side=left}{g1,t1}
	            \fmf{dashes,label=$p$,l.side=right}{g2,t1}
		    \fmf{dashes,label=$q$,l.side=right}{h1,t3}
		    \fmf{dashes,label=$q'$,l.side=left}{z1,t4}
		    \fmf{fermion,right=0.7,tension=0.7,label=$\psi$,l.side=right}{t1,t3}
		    \fmf{plain,right=0.3}{t3,t4}
		    \fmf{fermion,right=0.7,tension=0.7,label=$\bar{\psi}$,l.side=right}{t4,t1}
		    \fmf{arrow_right}{g1,t1}
		    \fmf{arrow_left}{g2,t1}
		    \fmf{arrow_right}{t3,h1}
		    \fmf{arrow_left}{t4,z1}
	        \end{fmfgraph*}
	    \end{fmffile}}
\end{center}
\caption{\label{fig:dmdmSS}\footnotesize
One-loop contributions to the DM-DM-$S$-$S$ coupling.}
\end{figure}
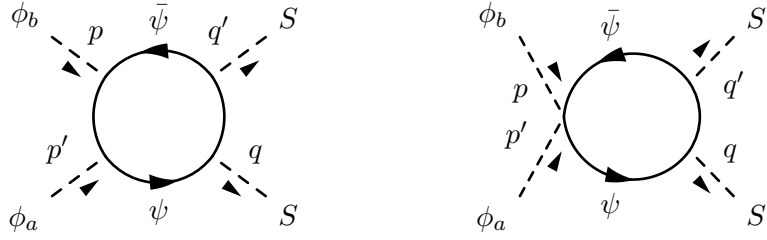

If the mass of scalar $S$ is sufficiently lighter than the DM mass, additional couplings shown in Fig.~\ref{fig:dmdmSS} will contribute to annihilation cross section. The four-point vertex function is given as
\begin{align}
\Gamma_{DM-DM-S-S}=&2 n_c y^2 \left( 1-\frac{G_D \langle\sigma\rangle}{8 G^2}\right)^2 \left(I_{6a}(p,p',q',M)+I_{6a}(p,p',q,M)\right) \nonumber \\
&+n_c y^2\frac{G_D}{4G^2}\left(I_{6b}(p,p',q',M)+I_{6b}(p,p',q,M)\right)
\end{align}
with the integrals given as
\begin{align}
I_{6a}(p,p',q',M)&=
\int \frac{d^4 k}{i(2\pi)^4}
\frac{\mbox{Tr}\gamma_5
(\slashed{k}+M)\gamma_5
(\slashed{k}-\slashed{p'}+M)
(\slashed{k}+\slashed{p'}-\slashed{q'}+M)
(\slashed{k}-\slashed{p}+M)}{(k^2-M^2)((k+p')^2-M^2)((k+p'-q')^2-M^2)((k-p)^2-M^2)}, \nn\\
I_{6b}(p,p',q',M)&=
\int \frac{d^4 k}{i(2\pi)^4}
\frac{\mbox{Tr}
(\slashed{k}+\slashed{p'}+M)
(\slashed{k}+\slashed{p'}-\slashed{q'}+M)
(\slashed{k}-\slashed{p}+M)}{((k+p')^2-M^2)((k+p'-q')^2-M^2)((k-p)^2-M^2)}.
\end{align}
This four-point function is only required when computing the relic abundance of DM, hence we only consider the case for $p=p'=(m_{\rm DM},\bf0)$ and denote the coupling as $\varkappa_s$. 

\subsection{Dark Matter Relic Abundance and its Direct Detection}

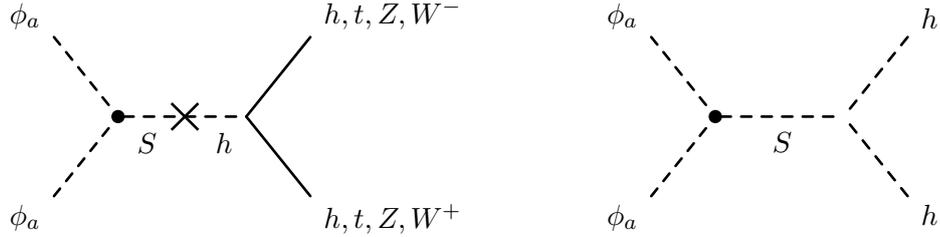
\begin{figure}[t]
\begin{center}
\subfloat{
\begin{fmffile}{schannel1}
	        \begin{fmfgraph*}(120,60)
		    \fmfcmd{vardef cross_bar (expr p,len,ang) = ((-len/2,0)--(len/2,0)) rotated (ang + angle direction length(p)/2 of p) shifted point length(p)/2 of p enddef;
		      style_def crossed expr p = draw_dashes p; ccutdraw cross_bar (p,5mm,45); ccutdraw cross_bar (p,5mm,-45) enddef;}  
	            \fmfleft{g1,g2}
		    \fmflabel{$\phi_a$}{g1}
                    \fmflabel{$\phi_a$}{g2}
	            \fmfright{h1,z1}
                    \fmflabel{$h,t,Z,W^+$}{h1}
		    \fmflabel{$h,t,Z,W^-$}{z1}
	            \fmf{dashes}{g1,t1}
	            \fmf{dashes}{g2,t1}
		    \fmf{crossed,label=$S\qquad h$}{t1,t2}
		    \fmf{plain}{h1,t2}
		    \fmf{plain}{z1,t2}
		    \fmfdot{t1}
	        \end{fmfgraph*}
	    \end{fmffile}}
\hspace{3cm}
\subfloat{
\begin{fmffile}{schannel2}
	        \begin{fmfgraph*}(120,60)
		    \fmfcmd{vardef cross_bar (expr p,len,ang) = ((-len/2,0)--(len/2,0)) rotated (ang + angle direction length(p)/2 of p) shifted point length(p)/2 of p enddef;
		      style_def crossed expr p = draw_dashes p; ccutdraw cross_bar (p,5mm,45); ccutdraw cross_bar (p,5mm,-45) enddef;}  
	            \fmfleft{g1,g2}
		    \fmflabel{$\phi_a$}{g1}
                    \fmflabel{$\phi_a$}{g2}
	            \fmfright{h1,z1}
                    \fmflabel{$h$}{h1}
		    \fmflabel{$h$}{z1}
	            \fmf{dashes}{g1,t1}
	            \fmf{dashes}{g2,t1}
		    \fmf{dashes,label=$S$}{t1,t2}
		    \fmf{dashes}{h1,t2}
		    \fmf{dashes}{z1,t2}
		    \fmfdot{t1}
	        \end{fmfgraph*}
	    \end{fmffile}}
\end{center}
\caption{\label{fig:dmdmSM}\footnotesize
Annihilation of DM into the SM particles.
The s-channel DM-DM-S coupling is $\kappa_s$, which is
given in  Eq.~(\ref{kappas}). }
\end{figure}

Now we are in position to compute the relic abundance of DM and its cross section with nuclei.
In Fig.~\ref{fig:dmdmSM} we show the  diagrams for DM annihilation into the SM particles.
The t-channel contributions are of $\mathcal{O}(y^4)$ due to two $\mathrm{DM-DM}-S$ coupling insertions, and
furthermore they are in higher order in $1/n_c$.
There is also a one-loop contribution
to the $\phi^2 S^2$ coupling as shown in 
Fig.~\ref{fig:dmdmSS}, which is also $\mathcal{O}(y^4)$.
These contributions would be important if 
the singlet $S$ is lighter than DM
for a relatively large $y$. We, however,
found that in this region of the parameter space we can not obtain
realistic values for $\Omega \hat{h}^2$.
Therefore, we will neglect these contributions and 
furthermore only take into account the s-wave contribution
to the s-channel annihilation cross sections, which are further enhanced by resonance effects.
We find that the s-wave contribution to
the thermal average $\langle v \sigma\rangle$ is given by
\be
\langle v \sigma\rangle
&=&\frac{{Z_{\rm DM}^2}}{32\pi m_{\rm DM}^3}
\left[ (m_{\rm DM}^2-M_W^2)^{1/2} a_W
+ (m_{\rm DM}^2-M_Z^2)^{1/2} a_Z \right.\nn\\
& &\left. + (m_{\rm DM}^2-M_t^2)^{3/2} a_t
 +  (m_{\rm DM}^2-m_h^2)^{1/2} a_h\right)+\mathcal{O}(v^2)~,
\ee
where {$Z_{\rm DM}$ is given in (\ref{wave0}), }
\be
a_W&=& 4 (\kappa_s/v_h)^2\left| \Delta_{hs}
\right|^2 M_W^4\left( 3+4\frac{m_{\rm DM}^4}{M_W^4}-4
\frac{m_{\rm DM}^2}{M_W^2}\right)~,\nn\\
a_Z&=& 2 (\kappa_s/v_h)^2\left| \Delta_{hs}
\right|^2  M_Z^4\left( 3+4\frac{m_{\rm DM}^4}{M_Z^4}-4
\frac{m_{\rm DM}^2}{M_Z^2}\right)~,\nn\\
a_t&=& 24 (\kappa_s/v_h)^2\left| \Delta_{hs}
\right|^2  m_t^2~,\nn\\
a_h &=&\frac{1}{2}(\kappa_s/v_h)^2(M_W/g)^2 
\left|~24 \lambda_H \Delta_{hs}
-4 \lambda_{HS}(v_s /v_h)\Delta_{ss}
\right|^2~,
\ee
 with $v_h=246$ GeV, and 
\be
\Delta_{hs} &=&\frac{\xi_2^{(2)}\xi_1^{(2)}}{4 m_{\rm DM}^2-m_S^2+i \gamma_S m_S}+
\frac{\xi_2^{(1)}\xi_1^{(1)}}{4 m_{\rm DM}^2-m_h^2}~,\nn\\
\Delta_{ss} &=&\frac{\xi_2^{(2)}\xi_2^{(2)}}{4 m_{\rm DM}^2-m_S^2+i \gamma_S m_S}+
\frac{\xi_2^{(1)}\xi_2^{(1)}}{4 m_{\rm DM}^2-m_h^2}~.
\ee
 Here $\kappa_s$ and $  \xi's$ are given in Eq.~(\ref{kappas}) and Eq.~(\ref{eigenstate}), respectively, $g\simeq 0.632$ is the $SU(2)_L$ gauge coupling,
and  
\begin{align}
\gamma_S= \frac{(\lambda_{HS}\langle S \rangle)^2}{8\pi m_S^2}\sqrt{\frac{m_S^2}{4}-m_h^2} 
\end{align}
is the decay width of $S$.

Given the annihilation cross {section} we can now compute
the relic abundance. To this end we use
the approximate formula \cite{Griest:1988ma}
\be
\Omega \hat{h}^2 &=&8\times 
\frac{Y_\infty s_0 m_{\rm DM}}{\rho_c/\hat{h}^2}~
\mbox{with}~Y_\infty^{-1} =
0.264 g_*^{1/2} M_{pl} m_{\rm DM}\langle v \sigma\rangle/x_f~,
\label{omega}
\ee
where $Y_\infty$ is the asymptotic value of the ratio $n_{\rm DM}/s$,
$s_0=2970/\mbox{cm}^3$ is the entropy density at present,
$\rho_c=3 H^2/8 \pi G=
1.05 \times 10^{-5}\hat{h}^2 ~\mbox{GeV}/\mbox{cm}^3$ is the critical density,
$\hat{h}$ is the dimensionless Hubble parameter,
 $M_{pl}=1.22\times 10^{19}~ \mbox{GeV}$ is the  Planck energy,
and $g_*=115.75$ is the number of the effectively massless degrees of freedom
at the freeze-out temperature.
Further, $x_f$ is the ratio $ m_{\rm DM}/T$ at the 
freeze-out temperature and can be obtained from
\cite{Griest:1988ma}
\be
x_f &=&\ln \frac{0.0764 M_{pl}
\langle v \sigma\rangle (5/4) m_{\rm DM}}
{(g_* x_f)^{1/2} }~.
\label{xf}
\ee
We multiplied with $8$ in (\ref{omega}), because 
there are $8$ DM particles.

We next come to the spin-independent 
elastic cross section off the nucleon 
$\sigma_{SI}$, which is shown in Fig. \ref{sigmaSI}
is given by \cite{Barbieri:2006dq}
\be
\sigma_{SI}
&=&\frac{{Z_{\rm DM}^2}}{\pi} 
\left[ \frac{\kappa_t \hat{f}
m_N }{2 v_h m_{\rm DM}}
\left(\frac{\xi_2^{(2)}\xi_1^{(2)}}{m_S^2}+
\frac{\xi_2^{(1)}\xi_1^{(1)}}{m_h^2}\right) \right]^2
\left(\frac{m_N m_{\rm DM}}{m_N+m_{\rm DM}}
\right)^2~,
\label{sigmaSI}
\ee
where $\kappa_t$ is given in
 (\ref{kappat}), $m_N$ is the nucleon mass, and
$\hat{f}\sim 0.3$ stems from the nucleonic matrix element 
\cite{Ellis:2000ds}.

\begin{figure}
  \includegraphics[width=5cm]{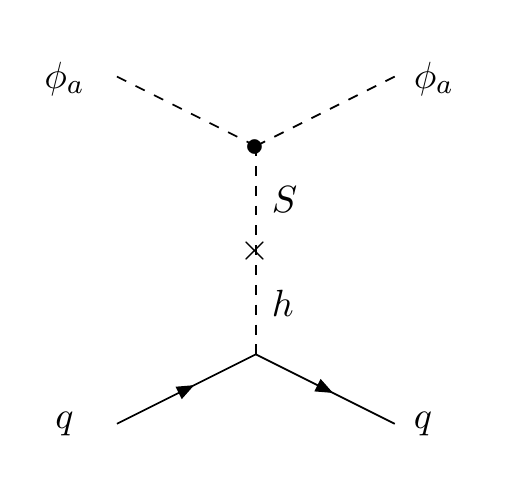}
\caption{\label{fig:sigmaSI}\footnotesize
Diagram contributing to the direct detection of DM.
The DM-DM-S coupling is $\kappa_t$, which is given
in (\ref{kappat}).}
\end{figure}

\begin{figure}
  \includegraphics[width=8cm]{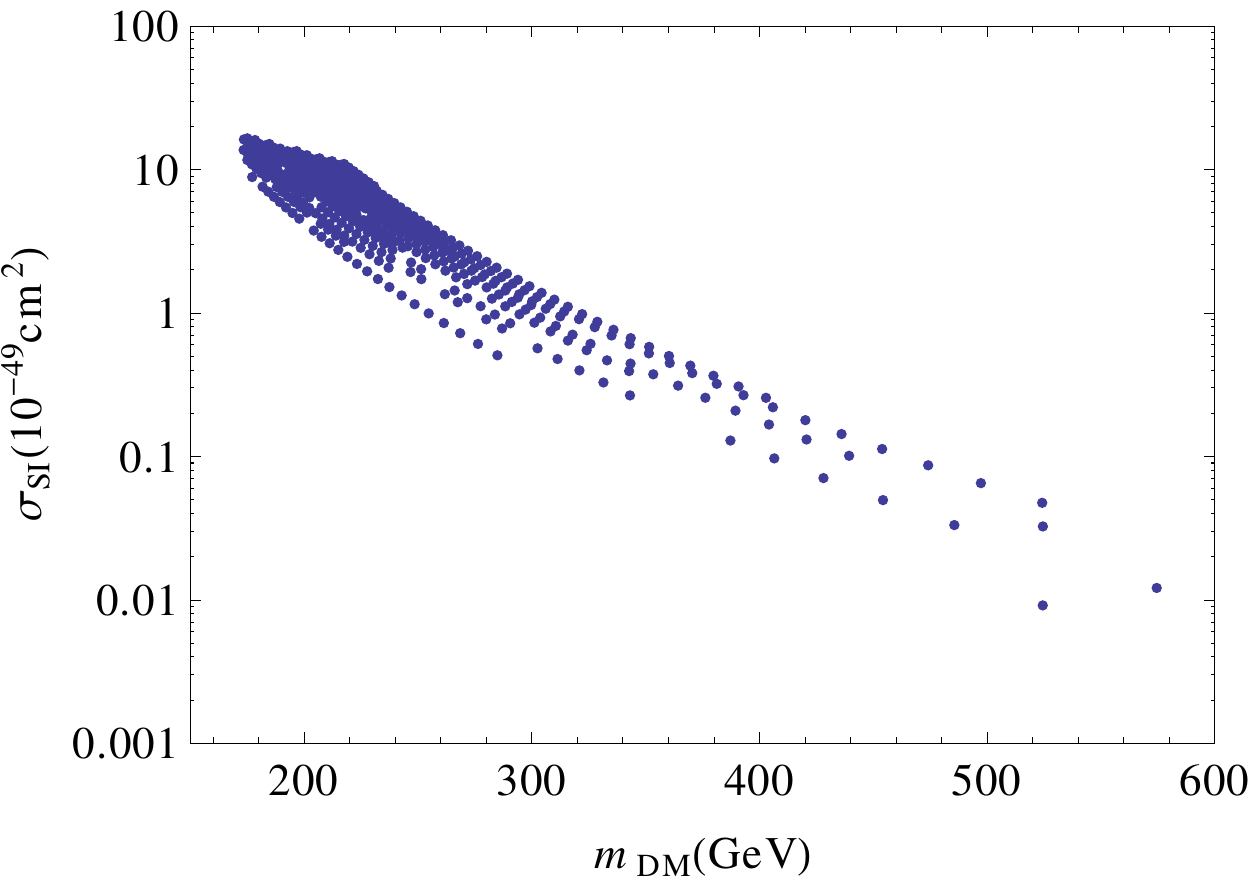}
\caption{\label{fig:mDM-sigma}\footnotesize
The DM mass $m_{\rm DM}$ against the 
spin-independent cross section $\sigma_{SI}$.
We have imposed:
$v_h=246~$ GeV, $m_h=125.9\pm 1.2$ GeV and $\Omega \hat{h}^2<
0.1187$. The XENON100 limit is
$3\times10^{-45}
~\mbox{cm}^2$ \cite{Aprile:2012nq},
while XENONIT will be sensitive down to $10^{-47}
~\mbox{cm}^2$  \cite{Aprile:2012zx}.}
\end{figure}

The constraints to be imposed are:
$v_h=\unit[246]{GeV}$, $m_h=\unit[125.9\pm 1.2]{GeV}$, $\Omega \hat{h}^2<
0.1187$, and $|\xi_1^{(1)}| \gtrsim  0.9$,
where these uncertainties correspond to $3\sigma$. We only assume that the relic abundance is less than the observed value as there could be another DM contribution such as the dark baryon. 
In Fig.~\ref{fig:mDM-sigma} we show in the  $m_{\rm DM}-\sigma_{SI}$ plane
the area in which all these constraints are satisfied.
Naively one may expect an extended area 
in the  $m_{\rm DM}-\sigma_{SI}$ plane, because we still have two free
parameters. But we see from Fig.~ \ref{fig:mDM-sigma} 
that the allowed area is
a narrow strip. This is because the coupling $\kappa_s$ is so small
that we have to use the resonant effect of the s-channel
diagrams in Fig.~\ref{fig:dmdmSM}. That is,  $2m_{\rm DM}\simeq 
m_S$ is required to obtain a realistic value of $\Omega \hat{h}^2$,
implying  that an extra  freedom is used in the parameter space. This model predicts no signal from the next generation direct DM detection experiments such as XENON1T and LUX. 
The parameter space of $\{ \lambda_H, \lambda_S, \lambda_{HS},y\}$
that can yield the allowed direct detection cross section and DM mass subjected by constraints above are given by
$\lambda_H \approx 0.13$, $\lambda_S \in (0.11,0.2)$, $\lambda_{HS} \in (0.001,0.05)$ and $y\in (0.003,0.007)$. We have also explicitly checked that $m_{\rm DM}<M$ such that the NJL method can be validly applied. This constraint has restricted the parameter space of $y$ in such a way that only $y$ of $\mathcal{O}(10^{-3})$ can reproduce the allowed relic abundance.

A simple extension of the model 
would be  to break the flavor group $SU(3)_V$ to a smaller
group by the Yukawa couplings as it is done in \cite{Hur:2007uz,Hur:2011sv}.
In doing so one may be able to relax the resonant constraint
$2 m_{\rm DM} \simeq m_S$, such that slightly extended area
in the $m_{\rm DM}-\sigma_{SI}$ plane is allowed. It is also possible to extend the model with another value of $n_f$ and $n_c$, the disadvantage of such an extension is that we are not allowed to scale up the known QCD values. However by changing $n_f$ or $n_c$, the NJL parameters $G$, $G_D$ and $\Lambda$ can be modified {and} it is possible to construct different  models.
Our model should be viewed as a prototype where we use the known values of QCD to demonstrate the general mechanism.

\section{Phase Transition at finite Temperature}\label{sec:phase transition}
We expect that at a certain finite temperature the chiral symmetry 
is restored \cite{Kirzhnits:1972ut}. Consequently, above that temperature 
the EW symmetry, too, must be restored.
The nature of the the EW symmetry breaking is intimately related to
Baryon asymmetry in the universe 
\cite{Kuzmin:1985mm,Klinkhamer:1984di,Arnold:1987mh,Shaposhnikov:1986jp}. Therefore it is interesting to test whether the model with allowed parameter space can yield a strong first order phase transition, which is a crucial ingredient for EW baryogenesis. We would like to investigate on how the chiral symmetry breaking and 
the EW symmetry breaking appear
as temperature decreases from a high temperature, which could play an important role in the thermal history of the universe.

\begin{figure}
\includegraphics[width=0.6\textwidth]{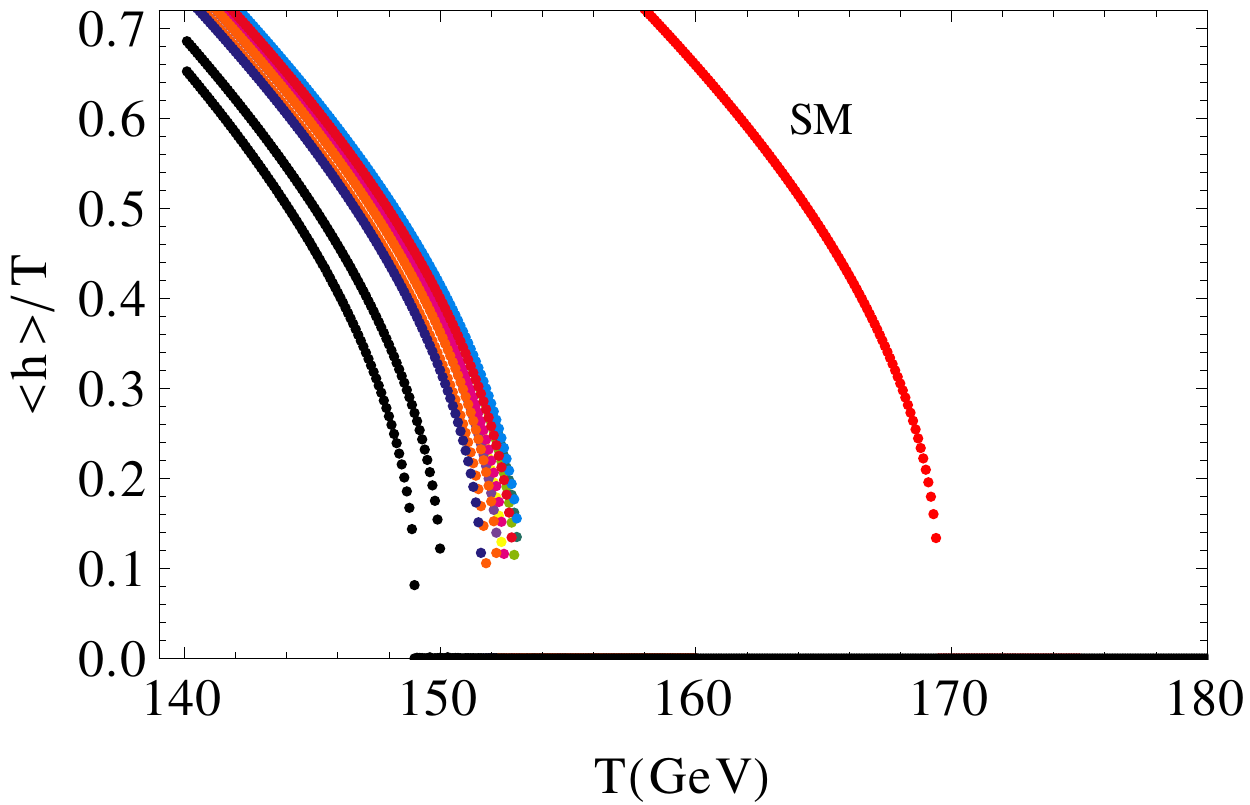}
\caption{\label{h-126-63}\footnotesize
The temperature dependence of 
$\langle h \rangle/T$ near the critical temperature
for the parameters used to obtain the points in Fig.~\ref{fig:mDM-sigma}.
The red points on the right side are for the SM. First order phase transition occurs around $T=\unit[150]{GeV}$.
}
\end{figure}

To answer the question on which order of phase transition, 
we will stay in NJL framework and integrate out the quantum
fluctuations at a finite temperature. As a result we obtain 
an effective potential at a finite temperature consists of five components \cite{Anderson:1991zb,Carrington:1991hz,Dine:1991ck,Dine:1992wr}
\footnote{The EW phase transition in the SM with singlets 
has been discussed in 
\cite{Kondo:1991jz,Benson:1993qx,Espinosa:1993bs,Choi:1993cv,Ham:2004cf,Ahriche:2007jp}. Note that in contrast to
these works there are only dimensionless couplings
in the tree-level potential (\ref{eq:VSM}) in the present model.}:
\begin{align}
V_{\rm EFF}(\phi_c,T) &=V_{\mathrm{SM}+S}(\phi_c)+V_{\rm NJL}(\phi_c)+
V_{\rm CW}(\phi_c) +V_{\rm FT}(\phi_c,T)+V_{\rm RING}(\phi_c,T)~, \displaybreak[0]
\label{VEFF}
\end{align}
where $\phi_c$ represents a collection of the classical scalar fields $h$, $S$ and $\sigma$. The term $V_{\mathrm{SM}+S}(\phi_c)$ is the tree-level
contribution given in Eq.~(\ref{eq:VSM}), $V_{\rm NJL}(\phi_c)$ as the one-loop effective potential (\ref{eq:Vnjl}) when the dark fermions are integrated out, $V_{\rm CW}(\phi_c)$ is the one-loop effective potential contribution for the rest of the fields at $T=0$, and $V_{\rm RING}$ is the ring contribution for the bosons. In the one-loop order 
they are given by, respectively,
\be
V_{\rm CW}(\phi_c) &=&\frac{1}{64\pi^2}\sum_i n_i \left\{ m_i^4(\phi_c)
\left(\ln \left[ \frac{m_i^2(\phi_c)}{m_i^2(\langle \phi_c \rangle)}\right] -\frac{3}{2}\right) +2m_i^2(\langle \phi_c \rangle)m_i^2(\phi_c) \right\}\label{VCW}~,\\
V_{\rm FT}(\phi_c,T)
&=&\frac{T^4}{2\pi^2}
\left(\sum_i n_i^BJ_B(m_i^2(\phi_c)/T^2)+
\sum_i n_i^FJ_F(m_i^2(\phi_c)/T^2)\right)\label{VFT}~,\\
V_{\rm RING}(\phi_c,T)
&=&-\frac{T}{12\pi}\sum_i n_i^B
\left[(M_i^2(\phi_c,T))^{3/2}-(m_i^2(\phi_c))^{3/2}\right]~,
\label{VRING}
\ee
where $n_i=n_i^B=1$ and $3$  for a real scalar and a vector boson, 
respectively, $n_i=n_i^F=-4$  for a Dirac fermion. Note that we include only the contribution from the top quark, the EW gauge bosons and the scalars $h$ and $S$ in the Coleman-Weinberg potential and the ring correction, as the contributions from Nambu-Goldstone bosons and the rest of the SM fermions are small. An additional contribution from the hidden constituent quark\footnote{
Integrating out the fermions in the hidden sector we obtain
the contribution to $V_{\rm FT}$:
$$V_{\rm NJL}(T,\sigma,S)=
-6n_c \frac{T^4}{\pi^2}J_F(M^2/T^2)
 \simeq
3n_c
\frac{T^2}{12}M^2+\frac{3n_c}{16 \pi^2}
\left[
M^4 \ln\left
( \frac{M^2}
{\pi^2 T^2 e^{3/2-\gamma_E}} \right)
\right], $$ where the current mass $M$ is given in (\ref{eq:cmass0}).} is also present in the $V_{\rm FT}$ potential. 
The tree level field dependent mass $m_i^2(\phi_c)$ and the thermal mass $M_i^2(\phi_c,T)$ for boson $i$ are given in Appendix \ref{sec:thermal}, and the thermal functions $J_B$ and $J_F$ are defined as
\be
J_B(r^2) &=&\int_0^\infty dx x^2 \ln
\left(1-e^{-\sqrt{x^2+r^2} } \right)\nn\\
&\simeq &
-\frac{\pi^4}{45}+\frac{\pi^2}{12}r^2
-\frac{\pi}{6}r^{3}-\frac{r^4}{32}
\left[\ln (r^2 /16\pi^2)+2\gamma_E-\frac{3}{2}   \right]+\dots~,
\label{JB}\\
J_F(r^2) &=&\int_0^\infty dx x^2 
\ln\left(1+e^{-\sqrt{x^2+r^2} } \right)\nn\\
&\simeq &
\frac{7\pi^4}{360}-\frac{\pi^2}{24}r^2
-\frac{r^4}{32}\left[\ln (r^2 /\pi^2)+2\gamma_E-\frac{3}{2}
   \right]+\dots, \label{JF}
\ee
where we have used the high temperature expansion to simplify our calculation. By using the same approximation we will determine the phase transition for the case of the SM as well. 

\begin{figure}
\includegraphics[width=0.445\textwidth]{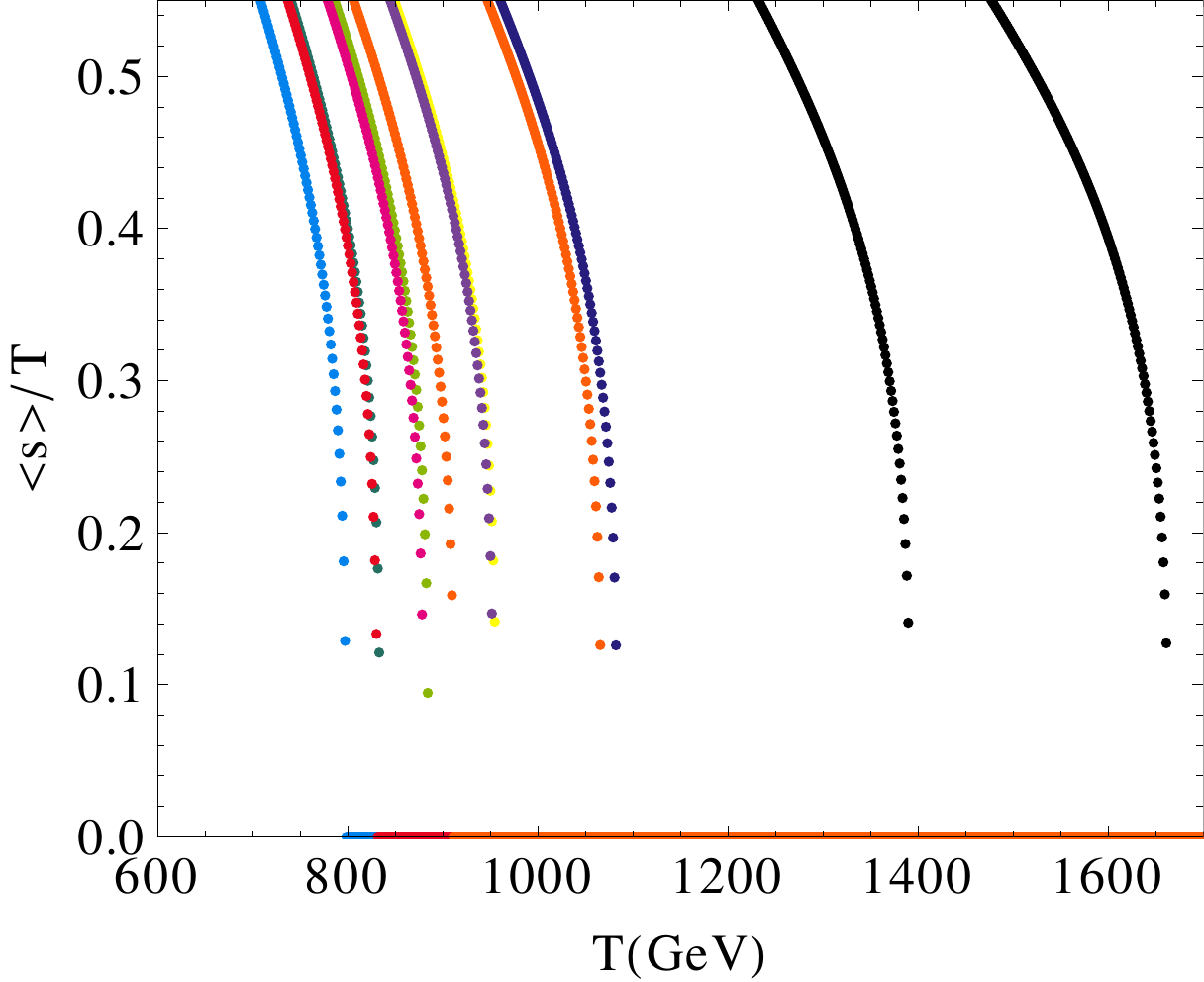}
\hspace{0.3cm}
\includegraphics[width=0.445\textwidth]{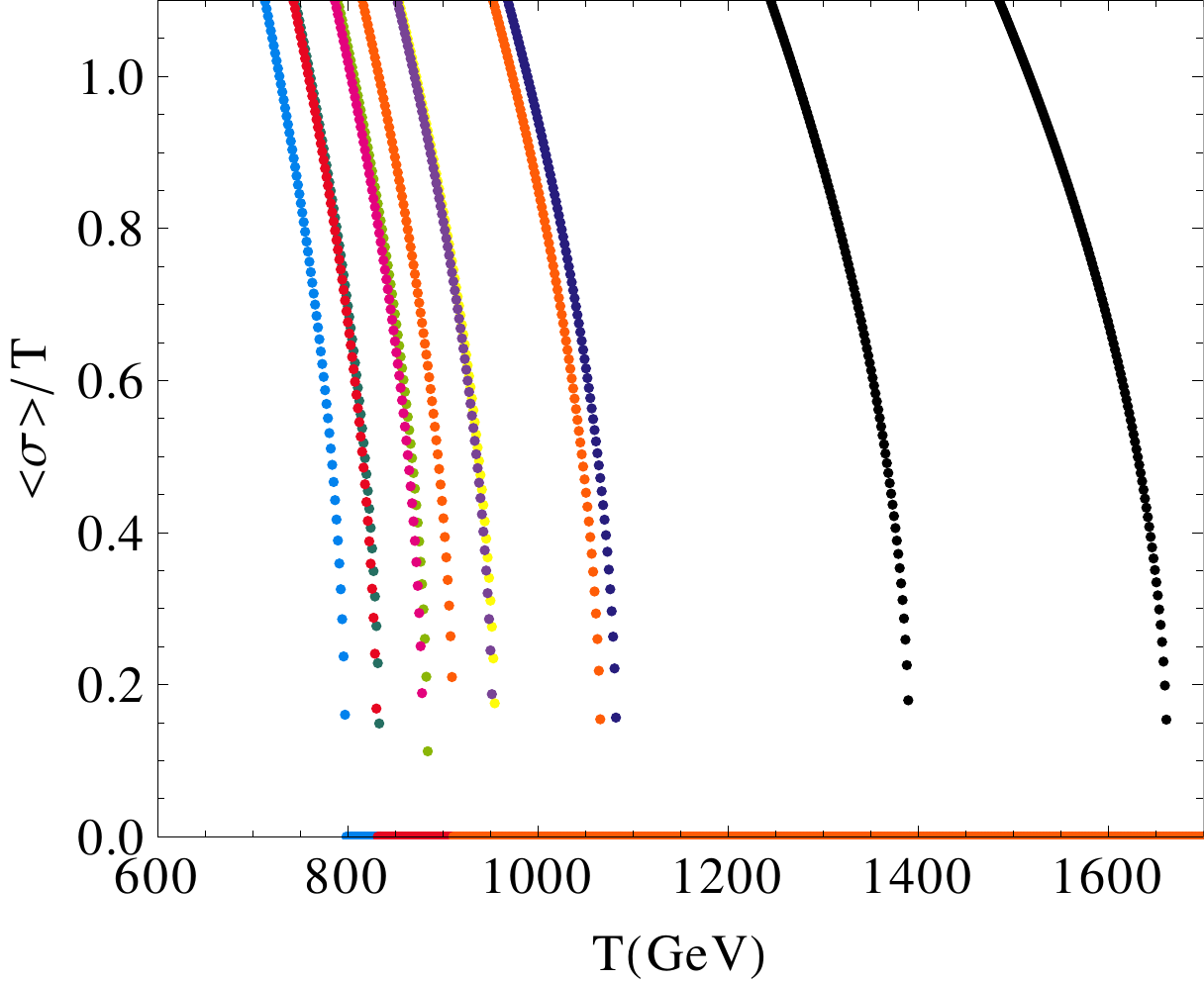}
\caption{\label{fig:ssigma}\footnotesize
The temperature dependence of 
$\langle S \rangle/T$ (left) and $\langle \sigma \rangle/T$ (right) near the critical temperature
for the parameters used to obtain the points in Fig.~\ref{fig:mDM-sigma}. First order phase transition takes place for our data set with critical temperature ranging from $T=\unit[800]{GeV}$ to $\unit[1700]{GeV}$.
}
\end{figure}

In Fig.~\ref{h-126-63} we show the temperature dependence of 
$\langle h \rangle/T$ near the critical temperature
for the parameter space that predicts acceptable relic abundance, i.e.~$\lambda_H \approx 0.13$, $\lambda_S \in (0.11,0.2)$, $\lambda_{HS} \in (0.001,0.05)$ and $y\in (0.003,0.007)$.
The red points are plotted for the case of the SM as reference. We sample a small amount of data to give  an idea where the phase transition occurs. The critical temperature is around $\unit[150]{GeV}$ for our data set. 
We see from this figure that the EW phase transition is of first order and that the critical temperature of the present model is always smaller than that of the SM. The shift of critical temperature from the SM is due to the non-negligible value of $\lambda_{HS}$. From Fig.~\ref{h-126-63} we can conclude from the model with the allowed parameter space predicts only weak first order EW phase transition, i.e. $\langle h \rangle /T_c <1$, therefore it cannot account for EW baryogenesis. However care must be taken when such a conclusion is drawn as we would also require non-perturbative calculation for a more accurate analysis.

\begin{figure}
\includegraphics[width=0.445\textwidth]{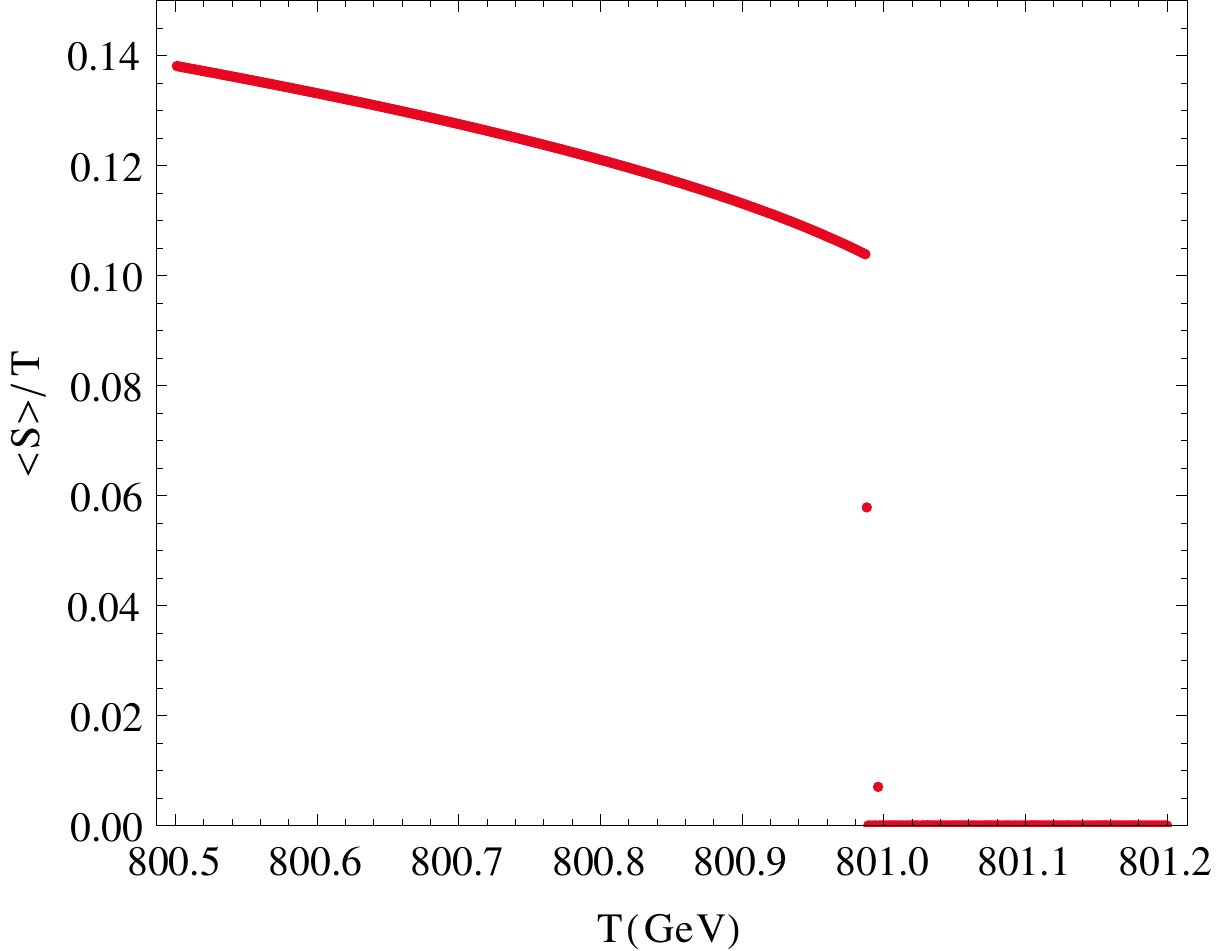}
\hspace{0.3cm}
\includegraphics[width=0.445\textwidth]{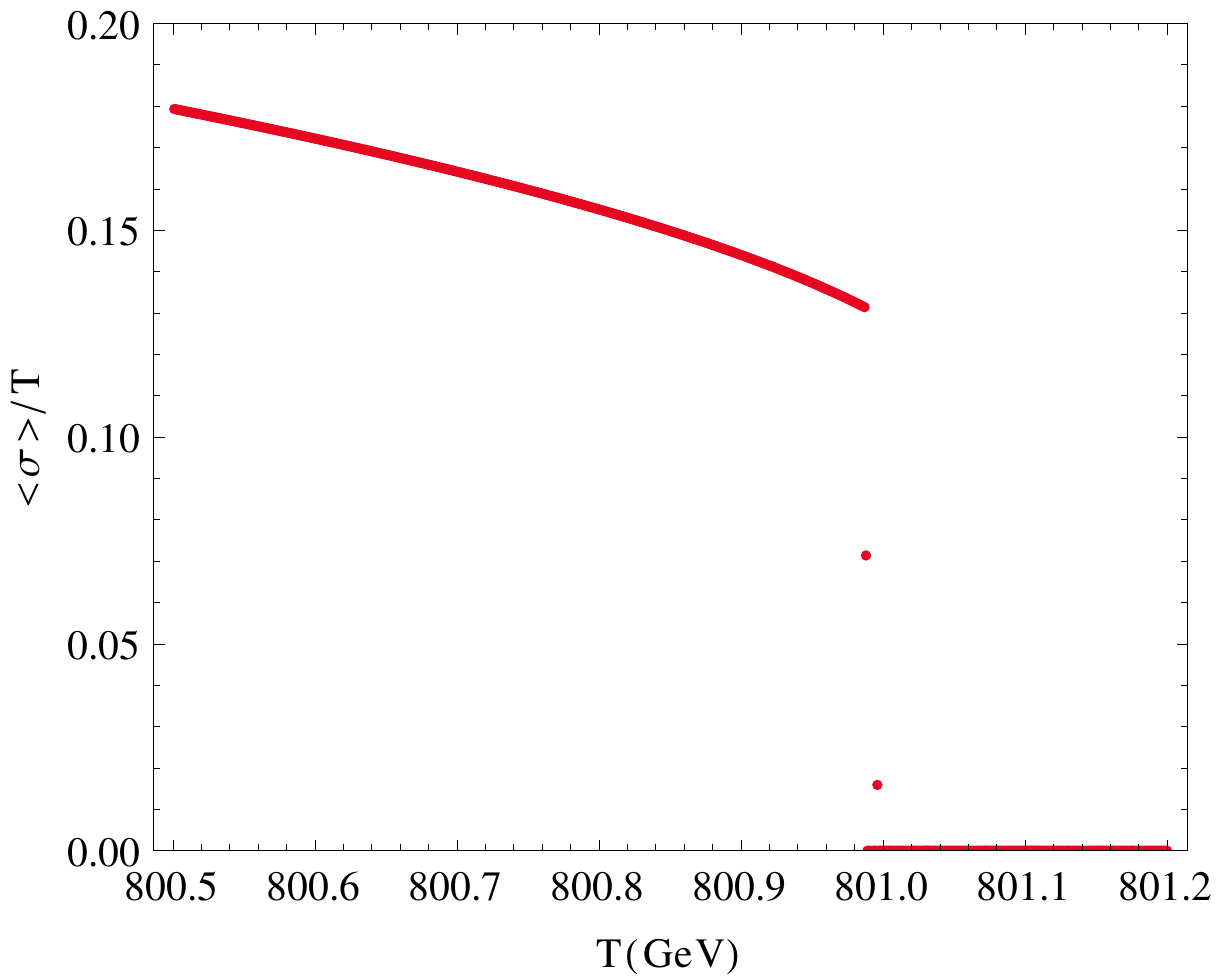}
\caption{\label{fig:ssigmazoom}\footnotesize
The temperature dependence of 
$\langle S \rangle/T$ (left) and $\langle \sigma \rangle/T$ (right) near the critical temperature $T_c\approx \unit[801]{GeV}$. The clear jump at critical temperature depicted in both diagrams indicates a weak first order phase transition.
}
\end{figure}

We now turn to the chiral phase transition in the dark sector and also the condensation of the real scalar mediator. We show the temperature dependence of $\langle S \rangle /T$ and $\langle \sigma \rangle /T$ in Fig.~\ref{fig:ssigma}. As we can see in both diagrams, the phase transition in the dark sector occurs from $T=\unit[800]{GeV}$ to $\unit[1700]{GeV}$ and all of them are of the first order type and hence bubble nucleation can possibly  occur during the thermal expansion of the universe. The zoom-in plots for an example curve near the critical temperature are shown in Fig.~\ref{fig:ssigmazoom} and we can conclude that the phase transition for the real scalar mediator and the chiral phase transition for the hidden sector are weakly first order. We would like to stress that our result is based on the NJL approach. A more accurate calculation based on lattice simulation or functional renormalization groups could alter the result.

\section{Summary and Conclusions}\label{sec:summary}
With no new signs of new physics from the LHC, our long-held believes on naturalness should be scrutinized. 
We argued in this paper that conformal symmetry might act as protective symmetry which could provide an alternative solution to the hierarchy problem. 
We have studied therefore a strongly coupled hidden sector, which we took to be a dark copy of QCD with unbroken flavor symmetry coupled to the SM via a singlet scalar.
The strongly coupled sector triggers a spontaneous chiral symmetry breaking and transmits the breaking scale to the SM sector. 
To describe the transmission not only qualitatively but also quantitatively, we have to be able to understand quantitatively the interplay between $ \langle S\rangle, \langle h\rangle$ and the chiral condensate in the hidden sector.
The NJL model provides us a appropriate framework to describe our low energy effective theory, where we made a nontrivial assumption that the values of the parameters of the NJL model corresponding to the hidden QCD are, up to the overall scale, the same as those values of the NJL model corresponding to QCD which describes the real hadron world.
Once this is accepted we have the same number of the free parameters as in the  hidden sector Lagrangian ${\cal L}_H$, which we can use to study the emergence of EW symmetry breaking scale from the dark sector.

The strongly coupled hidden sector provides naturally stable cold DM candidates, e.g. the dark pions that are massive because of the Yukawa coupling.
Needless to say that the DM-DM-S coupling, which is essential for DM analysis, can be directly computed in the NJL model, in contrast to the usual linear and non-linear sigma model.
It turned out that the thermally averaged cross-section is quite suppressed for most of the parameter space and it is therefore necessary to adjust parameters such that a resonance condition is fulfilled. This boosts the cross-section and suppresses the abundance sufficiently. Unfortunately this constraint of the allowed parameter regions implies that the DM nucleon cross-section is highly suppressed and there is therefore little prospect of direct detection of the DM candidate in next generation experiments.

We also used the NJL formalism to study phase transitions in the whole system, which possesses three order parameters. We found that both the EW phase transition and  the chiral phase transition are 
weakly first order, therefore EW Baryogenesis cannot be explained in this model.
Of course,  more accurate calculations based on lattice simulations,
for instance, are needed to confirm this observation.

The analysis performed here may be useful to study other models as well: for example, a straightforward extension is to make the singlet complex.
Then its CP odd part-an axion in the hidden sector- will be also a DM candidate,
which will mix with $\eta_0$ and will be lighter than the hidden pion DM.
We expect  a different DM phenomenology,
which will be the next target of our future project.
At last and not at least we
emphasize that models with different $n_c$ and $n_f$ is 
an interesting extension.

\vspace{0.5cm}
\noindent {\bf Acknowledgements:} We would like to thank Ken-Ichi Aoki, Daisuke Sato, Shinji Takeda for useful discussions.
J.~K. would like to thank the theory group of the Max-Planck-Institut f\"ur Kernphysik in Heidelberg for their hospitality.
J.~K. is partially supported by the Grant-in-Aid for Scientific Research (C) from the Japan Society for Promotion of Science (Grant No.22540271).
K.~S.~L. acknowledges support by the International Max Planck Research School for Precision Tests of Fundamental Symmetries. 
\appendix

\section{Determination of the NJL parameters \mbox{\texorpdfstring{$G,\,G_D$}{G,GD}} and \mbox{\texorpdfstring{$\Lambda$}{Lambda}} \label{sec:appendix}}
\subsection{The free part \mbox{\texorpdfstring{${\cal L}_0$}{L0}} in the SCMF approximation}
Here we give a more detailed description of the NJL formalism
\cite{Kunihiro:1983ej,Kunihiro:1987bb,Hatsuda:1994pi}.

The interaction part, $2G~\mbox{Tr} ~\Phi^\dag \Phi$ 
in (\ref{eq:NJL10}), can 
be written as
\be
2G~\mbox{Tr} ~\Phi^\dag \Phi&=&G\sum_{a=0}^8\left[
(\bar{\psi}\lambda^a \psi)^2+ (i\bar{\psi}\gamma_5\lambda^a \psi)^2
\right]~,
\label{eq:NJLint}
\ee
and we can write the second term of 
the rhs of (\ref{eq:NJLint}) as
\begin{align}
G\sum_{a=0}^8(i\bar{\psi}\gamma_5\lambda^a \psi)^2&=
G\sum_{a=0}^8(i\bar{\psi}\gamma_5\lambda^a \psi
+\frac{1}{2G}\phi_a)^2 -\frac{1}{4G}\sum_{a=0}^8 \phi_a\phi_a
 -\sum_{a=0}^8 i\bar{\psi}\gamma_5\lambda^a \psi
~\phi_a~.
\label{eq:phi}
\end{align}
We then regard the first term of (\ref{eq:phi}) as an interaction term
and according to the SCMF approximation
\cite{Kunihiro:1983ej,Kunihiro:1987bb,Hatsuda:1994pi}
we rewrite it as normal products:
\begin{align}
G\sum_{a=0}^8(i\bar{\psi}\gamma_5\lambda^a \psi 
+\frac{1}{2G}\phi_a)^2=&
G\sum_{a=0}^8 :(i\bar{\psi}\gamma_5\lambda^a \psi )^2:+G\sum_{a=0}^8 (
i\widehat{\bar{\psi}\gamma_5\lambda^a \psi}  
+\frac{1}{2G}\phi_a  )^2\nn\\
&+2G\sum_{a=0}^8 : i\bar{\psi}\gamma_5\lambda^a \psi :
(i\widehat{\bar{\psi}\gamma_5\lambda^a \psi}  
+\frac{1}{2G}\phi_a  )~.
\label{eq:normal}
\end{align}
The normal product and contraction denoted by 
$\widehat{~~~}$ are defined with respect to the vacuum
of the fermion, where the vacuum is defined
by the fermion bi-linear  part of the Lagrangian,
which we will denote by ${\cal L}_0$.
Further, the last two terms in (\ref{eq:normal}) vanish if we identify the meson field as in Eq.~(\ref{eq:phi1}). 
This identification of the meson field and the definition of the vacuum
is known as bosonization and it is the essential part of the SCMF approximation.
For the scalar part
we rewrite it  in a similar way.

The anomaly term can also be treated in a similar manner.
Using the result of Cayley-Hamilton theorem
\be
\det \Phi &=&
\frac{1}{3}\mbox{Tr}~ \Phi^3
-\frac{1}{2}\mbox{Tr}~ \Phi^2~\mbox{Tr}~ \Phi
+\frac{1}{6}(\mbox{Tr}~ \Phi)^3~,
\ee
we find
\be
G_D\left(\mbox{Tr} \varphi^2\Phi
-\mbox{Tr} \varphi\Phi~\mbox{Tr} \varphi
-\frac{1}{2}
\mbox{Tr}\varphi^2\mbox{Tr} \Phi
+\frac{1}{2}(\mbox{Tr} \varphi)^2\mbox{Tr} \Phi +h.c.\right)-
2G_D (\det \varphi +h.c.)
\ee
should be added to the ``free''  part $ {\cal L}_0$.
Adding all together  we obtain
\be
{\cal L}_{\rm NJL}={\cal L}_0 +{\cal L}_I~,
 \ee
where
\begin{align}
{\cal L}_0 & =\mbox{Tr}\bar{\psi}(i\gamma^\mu\partial_\mu -
y S)\psi +2G\mbox{Tr}(\varphi^\dag\Phi+h.c)
-2G\mbox{Tr}\varphi^\dag\varphi-
2G_D (\det \varphi +h.c.)
\nn\\
 &+G_D\left(\mbox{Tr} \varphi^2\Phi
-\mbox{Tr} \varphi\Phi~\mbox{Tr} \varphi
-\frac{1}{2}
\mbox{Tr}\varphi^2\mbox{Tr} \Phi
+\frac{1}{2}(\mbox{Tr} \varphi)^2\mbox{Tr} \Phi +h.c.\right)~,
\label{eq:L00}
\intertext{which can be simplified to (\ref{eq:L0}), and}
{\cal L}_I & =2G:\mbox{Tr}\Phi^\dag\Phi:+G_D:(\det \Phi+h.c):\nn\\
& +G_D:\left(\mbox{Tr} \varphi\Phi^2
-\mbox{Tr} \varphi\Phi~\mbox{Tr} \Phi
-\frac{1}{2}
\mbox{Tr}\Phi^2\mbox{Tr} \varphi
+\frac{1}{2}(\mbox{Tr} \Phi)^2\mbox{Tr} \varphi +h.c.\right):~,
\label{eq:LI}
\end{align}
which fulfills $\langle 0\vert \mathcal{L}_{I}\vert 0\rangle =0$, as required. 

As we assume scaled up values of $G$, $G_D$ and $\Lambda$ from their QCD values
\be
G= f^{-2} G^{\mathrm{QCD}}~,~G_D=  f^{-5} G_D^{\mathrm{QCD}}~,~\Lambda= f\Lambda^{\mathrm{QCD}}~,
\ee
 we first need to obtain their values from QCD. The following analyses had been performed in past by
Hatsuda and Kunihiro \cite{Kunihiro:1987bb,Hatsuda:1994pi}, who
 used a three-dimensional momentum cutoff.
To maintain  Lorentz covariance we have used a  four-dimensional momentum cutoff $\Lambda$, therefore our values vary slightly from Ref.~\cite{Kunihiro:1987bb,Hatsuda:1994pi}.
We summarize below the results for our case.

In the real QCD case the vector-like symmetry  $SU(3)_V$
is broken down to $SU(2)_V$ explicitly by the current quark masses,
which we denote by $m_1=m_2$ and $m_3$, and 
the singlet $S$ is absent.
Therefore, instead of Eq.~(\ref{eq:varphi}) we have
\begin{align}
 \widehat{\Phi} &=\widehat{\bar{\psi}_i\psi_j}-\widehat{\bar{\psi}_i\gamma_5\psi_j}=
 \varphi=
-\frac{1}{4G^{\mathrm{QCD}}} 
 \left(\mbox{diag.}(b,b,c)+i(\lambda^a)^T \phi^a\right)~.
\label{eq:varphi1}
\end{align}
To obtain $G^{\mathrm{QCD}}$, $G_D^{\mathrm{QCD}}$ and $\Lambda^{\mathrm{QCD}}$ we simply need to derive the NJL QCD Lagrangian in SCMF approximation and perform a fit from the calculated meson mass spectrum and the pion decay constant.

The derivation of the free part 
in the SCMF approximation is outlined above.
So, here we give the result for the case
that $SU(3)_V$ is explicitly broken by the current fermion masses,
where 
 $G$, $G_D$ and $\Lambda$ in the following equations mean
$G^{\mathrm{QCD}}$, $G_D^{\mathrm{QCD}}$ and $\Lambda^{\mathrm{QCD}}$, respectively:
\be
{\cal L}_0^{\mathrm{QCD}}& =&
i\mbox{Tr}\bar{\psi}\gamma^\mu\partial_\mu\psi-
\left(m_1+b-\frac{G_D}{8G^2}b c\right)\,\mbox{tr}\bar{\psi}\psi-
\left(m_3+c-\frac{G_D}{8G^2}b^2\right)\,\bar{\psi}_3\psi_3 -i \mbox{Tr}\bar{\psi}\gamma_5 \phi\psi\nn\\
& &+\frac{G_D}{8G^2}\left(
- \mbox{Tr}\bar{\psi}\phi^2 \psi+\sum_{a=1}^8\phi_a\phi_a \mbox{Tr}\bar{\psi}\psi +i c  ~\mbox{Tr}\bar{\psi}\gamma_5 \Pi \psi
+i b  ~\mbox{Tr}\bar{\psi}\gamma_5 K \psi \right)\nn\\
& & -\frac{1}{8G}\left(2b^2+c^2+2\sum_{a=1}^8\phi_a\phi_a\right)+\frac{G_D}{16G^3}\left(b^2 c+
c \sum_{a=1}^3(\Pi_a)^2
+b \sum_{a=1}^4(K_a)^2\right)~,
\label{eq:L01}
\ee
where we have defined
\be
\mbox{Tr}M &\equiv&\sum_{a=1,2,3}M_{a a}~,~\mbox{tr}M\equiv
\sum_{a=1,2}M_{a a}~,\nn\\
\phi &\equiv &
\sum_{a=1}^8 \equiv \lambda^a \phi_a,~\Pi_{1,2,3}\equiv \phi_i\lambda^i |_{1,2,3},~
K_{1,2,3,4}\equiv \phi_i\lambda^i |_{1,2,3,4}.
\ee
Note that the $\eta$ terms are omitted. 

\subsection{One-loop effective potential}
Once the QCD Lagrangian is known, the vacuum state can be obtained from the one-loop effective potential, which can be obtained by
integrating out the fermion fields:
\be
V_{\rm eff}(b,c)= \frac{1}{8G}(2b^2+c^2)-
\frac{G_D}{16G^3}(b^2 c)-2 n_c I_0(M_1)- n_cI_0(M_3)~,
\label{eq:Veff}
\ee
where 
the constituent masses $M_1$ and $M_3$ are, respectively, given by
\be
& & M_1 = m_1+b-\frac{G_D}{8G^2}b c,
M_3=m_3+c-\frac{G_D}{8G^2}b^2~,\label{eq:cmass}
\ee
and 
\begin{align}
I_0(m_0)&=\int\frac{d^4 k}{i(2\pi)^4}\ln \det (\slashed{k}-m_0) \nn \\ 
&=\frac{1}{16\pi^2}\left(\Lambda^4 \ln\left(1+ \frac{m_0^2}{\Lambda^2} \right)
-m_0^4 \ln\left( 1+\frac{\Lambda^2}{m_0^2} \right)
+m_0^2 \Lambda^2\right)~.
\label{eq:I0}
\end{align}
where $\Lambda$ is a four-dimensional momentum cutoff and $b$ and $c$ are defined in (\ref{eq:varphi1}).

\subsection{Meson mass and Pion decay constant}
The meson mass can be obtained from the zero of the
corresponding inverse propagator.
For the pion and Kaon we find
\begin{align}
\Gamma_\pi(p^2)
&=-\frac{1}{2G}+\frac{G_D}{8G^3}c
-(1-\frac{G_D}{8G^2}c)^2~2 n_c I_1(p^2,M_1)
+\frac{G_D}{G^2}n_c I_2(M_3)~, \nn \\
\Gamma_K(p^2)&=-\frac{1}{2G}+\frac{G_D}{8G^3}b
-(1-\frac{G_D}{8G^2}b)^2~n_c
\left(I_1(p^2,M_1)+I_1(p^2,M_3)\right)+\frac{G_D}{G^2}n_c I_2(M_1)~,
\label{sigma-pion}
\end{align}
where
\begin{align}
I_1(p^2,M)&=
\int \frac{d^4 k}{i(2\pi)^4}
\frac{\mbox{Tr}(k\hspace{-0.2cm}/-p\hspace{-0.2cm}/+M)
\gamma_5(k\hspace{-0.2cm}/+M)\gamma_5}{((k-p)^2-M^2)(k^2-M^2)},\nn \\
I_2(M)&=\int \frac{d^4 k}{i(2\pi)^4}
\frac{M}{(k^2-M^2)}
= -\frac{1}{16\pi^2}M\left[\Lambda^2-M^2 \ln\left(1+\frac{\Lambda^2}{M^2} \right)\right]~.
\label{I2}
\end{align}
The meson masses are  the zeros of the inverse propagators:
\be
\Gamma_\Pi(p^2=m_\pi^2)&=&0~,~\Gamma_K(p^2=m_K^2)=0~.
\label{zeros1}
\ee

The pion decay constant is defined as
\be
\langle 0| \mbox{Tr}\bar{\psi}\gamma_\mu\gamma_5
\frac{\sigma_a}{2}\psi|\Pi_b(p) \rangle
=i\delta_{ab}f_\pi p_\mu~.
\ee
The one-loop expression is given by
\be
f_\pi &=& Z_\pi^{1/2}n_c (1-\frac{G_D}{8G^2}c) I_3(m_\pi^2,M_1)~,
\label{decay}
\ee
where 
\be
 & &Z_\pi^{-1} =\left.
\frac{d \Sigma_\Pi (p^2)}{d p^2}\right|_{p^2=m_\pi^2}~,
\label{wave}\\
& & p_\mu I_3(p^2,M)=
\int \frac{d^4 k}{i(2\pi)^4}
\frac{\mbox{Tr}\gamma_\mu\gamma_5(k\hspace{-0.2cm}/-p\hspace{-0.2cm}/+M)
\gamma_5(k\hspace{-0.2cm}/+M)}{((k-p)^2-M^2)(k^2-M^2)}.\label{I3}
\ee

\subsection{Determination of the parameters}
The independent parameters in NJL QCD are:
\begin{align}
G^{\mathrm{QCD}},\;G_D^{\mathrm{QCD}},\;\Lambda^{\mathrm{QCD}},\;m_1,\;m_3.
\end{align}
It turns out that these five parameters can be fixed
from three   physical  quantities;
the pion mass $m_\pi$, the Kaon mass $m_K$ and
the pion decay constant $f_\pi$.
The best fit values of the parameters are given in Table \ref{tab:njlvalues}
together with  other quantities.
\begin{table}
\begin{tabular}{|c|c|c|c|c|c|c|c|c|} 
\hline
Parameter & $(2G^{\mathrm{QCD}})^{-1/2}$ & $(-G_D^{\mathrm{QCD}})^{-1/5}$ & $\Lambda^{\mathrm{QCD}}$  & $m_1$ & $m_3$& $m_{\pi}$ & $f_\pi$ & $m_K$ \\ \hline
Value (MeV) &326 &437 &924 & ${6.6}$ & 127& 138 &93&496 \\ \hline
\end{tabular}
\caption{Values of NJL QCD obtained by fitting the pion decay constant and the mass of pion and Kaon.}
\label{tab:njlvalues}
\end{table}

\section{Field dependent masses and thermal masses for bosons\label{sec:thermal}}
The tree level field dependent masses for relevant particles are given:
\begin{align}
m_W^2(h)&=\frac{g_2^2}{4}h^2,\quad m_Z^2(h)=\frac{g_2^2+g_1^2}{4}h^2,\quad m_t^2(h)=\frac{y_t^2}{2}h^2,
\end{align}
while the masses for $m_h^2(h,S)$ and $m_S^2(h,S)$ are given in Eq.~\eqref{eq:gammas}.

We calculate the relevant thermal masses $M_i(\phi_c,T)$ in Landau gauge and the values are given as follows:
\begin{align}
M^2_{W_L}(h,T)&=m_W^2(h)+\frac{11}{6}g_2^2 T^2, \nn \\
M^2_{Z_L}(h,T)&=m_Z^2(h)+\frac{11}{6}g_2^2 T^2, \nn \\
M^2_{hh}(h,S,T)&=m^2_{hh}(h,S)+\left( \frac{3g^2_2}{8}+\frac{\lambda_H}{2}+\frac{y_t^2}{4}-\frac{\lambda_{HS}}{24}\right)T^2, \nn \\
M^2_{SS}(h,S,T)&=m^2_{SS}(h,S)+\left( \frac{\lambda_S}{4}-\frac{\lambda_{HS}}{6}\right)T^2, \nn \\
M^2_{hS}(h,S,T)&\approx m^2_{hS}(h,S).
\end{align}
We ignore the contribution from Nambu-Goldstone bosons and $\mathrm{U}(1)_Y$ as they are small. The mass eigenstate for ${M^2_{hh},M^2_{SS}}$ is given by
\begin{align}
M^2_{1,2}(h,S,T)\approx & \frac{1}{2}\left( M^2_{hh}+M^2_{SS} \mp \sqrt{(M_{hh}^2-M_{SS}^2)^2-4 M^2_{hS}}\right). 
\end{align}
Note that only the longitudinal part of the EW gauge bosons contribute to their thermal masses.

\bibliographystyle{apsrev4-1}
\bibliography{strongly}
\end{document}